\documentclass{appolb}
\usepackage[dvips]{graphicx}% For importing graphics: .ps, .jpg, .bmp, .pcx, &
\usepackage{amsmath}
\usepackage{amsfonts}
\usepackage{amssymb}
\usepackage{longtable} % For tables that are too long for 1 page.
\usepackage{color}

\allowdisplaybreaks[4] % allows page breaks in the middle of an equation
\newcommand{\df}{\ {\overset {\rm def} =}\ }
\newcommand{\dr}[2]{\frac {{\rm d} {#1}} {{\rm d} {#2}}}
\newcommand{\pdr}[2]{\frac {\partial {#1}} {\partial {#2}}}
\newcommand{\dril}[2]{{{\rm d} {#1}} / {{\rm d} {#2}}}
\newcommand{\pdril}[2]{{\partial {#1}} / {\partial {#2}}}

\begin{document}
% \eqsec  % uncomment this line to get equations numbered by (sec.num)
\title{Gamma radiation from areal radius minima in a quasi-spherical
Szekeres metric}
\author{Andrzej Krasi\'nski
\address{N. Copernicus Astronomical Centre, Polish Academy of Sciences, \\
Bartycka 18, 00 716 Warszawa, Poland} \\
email: akr@camk.edu.pl}
 \maketitle
\begin{abstract}
In previous papers it was shown that in a quasi-spherical Szekeres (QSS) metric,
impulses of gamma radiation can arise that have several properties in common
with the observed gamma-ray bursts. This happens when the bang-time function
$t_B(r)$ has a gate-shaped hump around the origin of the QSS region. The gamma
rays arise along two preferred directions of the QSS geometry (coincident with
dipole extrema when axially symmetric, otherwise unrelated). In these
directions, the rays of the relic radiation are blueshifted rather than
redshifted. The blueshift is generated in a thin region between the Big Bang
(BB) and the extremum-redshift hypersurface (ERH). But the Szekeres models can
describe the real Universe only forward in time from the last-scattering
hypersurface (LSH) because the matter in them has zero pressure. The ERH is
tangent to the BB at the origin, so in a neighbourhood thereof the ERH lies
earlier than the LSH and no blueshift is generated in the physical region. The
question thus arose whether the BB and ERH can be ``unglued'' if the QSS region
has no origin, but the areal radius function $\Phi$ has a local maximum or
minimum somewhere. In the present paper it is demonstrated that this is indeed
the case. If the hump in $t_B(r)$ is centered around the minimum of $\Phi$, then
the BB and ERH in general do not coincide there and a stronger blueshift is
generated on rays passing nearby. It follows that a lower and narrower hump on
the BB set can generate sufficient blueshift to move the initial frequencies of
the relic radiation to the gamma range. These facts are demonstrated by
numerical calculations in an explicit example of a QSS region.
\end{abstract}
\PACS{PACS numbers come here}

\section{Motivation and background}\label{intro}

In previous papers \cite{Kras2016a} -- \cite{Kras2018b} it was shown that
flashes of gamma radiation with characteristics similar to those of the
gamma-ray bursts (GRBs) \cite{Cenk2011} -- \cite{BATSE} may arise in a
Lema\^{\i}tre \cite{Lema1933} -- Tolman \cite{Tolm1934} (L--T) and a
quasi-spherical Szekeres (QSS) model \cite{Szek1975,Szek1975b} if the Big Bang
(BB) function $t_B(r)$ has a suitably chosen profile in some regions. The
complete model of the Universe consisted of an L--T or QSS region embedded in a
$k < 0$ Friedmann background; each inhomogeneous region contained an origin
\cite{HeKr2002,PlKr2006}. The gamma radiation arises by blueshifting
\cite{Szek1980,HeLa1984} the light emitted at the end of the last scattering
epoch along radial directions in an L--T region \cite{Kras2016a} and along two
preferred directions in a QSS region \cite{Kras2016b,Kras2018a}.\footnote{These
preferred directions are in general unrelated to the mass-dipole axes
\cite{DeSo1985}, but coincide with them in an axially symmetric QSS model
\cite{Kras2016b}.} In Refs. \cite{Kras2016a,Kras2018a,Kras2018b} it was shown
that in this way one can imitate most observed properties of the GRBs: their
frequency (i.e. energy) range, the presence of afterglows, the collimation into
narrow jets, the large distances to their sources, the brief durations of the
bursts and their large number. However, two properties were in quantitative
disagreement with the observations: the durations of the afterglows in the
models were much longer than observed, and the angular radii of the sources seen
by a present observer (equal to $\approx 1^{\circ}$) were larger than the
localisation errors for most of the GRBs; see the last paragraph of this section
for an update on this.

Refs.  \cite{Kras2016a} -- \cite{Kras2018b} employed models in which the energy
function  had the Friedmannian form $E = - \tfrac 1 2 k r^2$ (with $k = - 0.4$)
everywhere, while the BB function $t_B(r)$ had a gate-shaped hump around the
origin of the inhomogeneous region. The blueshift is generated in a thin slice
of spacetime between the nonconstant segment of $t_B(r)$
\cite{Kras2016a,Kras2018a} and the etremum-redshift hypersurface (ERH) in L--T
models or extremum-redshift surface (ERS) in QSS models.\footnote{The numerical
prescription for determining the ERH is known only for L--T models
\cite{Kras2016a}. Numerical calculations imply that its analogue exists also in
QSS models \cite{Kras2018a,Kras2018b}, but an operational definition was found
only for the 2-dimensional ERS determined by rays proceeding along the symmetry
axis in axially symmetric QSS models.} After crossing the ERH/ERS to the future,
each ray acquires only redshift. If it were possible to observe rays emitted at
the BB, and if the real Universe had the L--T or QSS geometry down to the BB,
then all the blueshifted rays would display infinite blueshift to any later
observer, i.e. the observed radiation would have infinite frequency. The
redshift acquired after crossing the ERH/ERS could not compensate it.

However, the L--T and Szekeres models have zero pressure, so they do not apply
to the earliest cosmic epochs. It is assumed that they apply toward the future
from the last-scattering hypersurface (LSH) \cite{Kras2016a,Kras2018a}. (The LSH
in inhomogeneous models is that on which the local mass density is equal to
$\rho_{\rm LS}$ -- the mass density at last scattering in the $\Lambda$CDM
model, see Ref. \cite{Kras2016a} for the calculation and Eq. (\ref{2.15}) here
for the value.) The blueshift generated between the LSH and the ERH/ERS is
finite, and the redshift acquired later may overcompensate it. To generate a
strong blueshift, the hump on $t_B(r)$ should be sufficiently high and wide, but
to keep the perturbations of the cosmic microwave background radiation within
the limits allowed by observations, the hump should be as low and narrow as
possible. These contradictive factors must be balanced to ensure that the
initially generated blueshift is strong enough to survive the later redshifting
while the height and diameter of the hump on $t_B(r)$ are within tolerable
limits.

In Refs. \cite{Kras2016a} -- \cite{Kras2018b} the hump on $t_B(r)$ was centered
on the origin of the L--T or QSS region (at $r = 0$ in the coordinates used
there), where the ERH or ERS was tangent to the BB \cite{Kras2016a,Kras2018a}.
Then, the blueshift-generating region disappears at $r = 0$ and is thin (in the
timelike direction) in a neighbourhood. The question thus arose whether the
ERH/ERS and the BB could be ``unglued''. It is shown in Appendix \ref{ERSatorig}
that, if the hump on $t_B(r)$ is centered at the origin, then the unglueing of
ERS and BB is possible only at the cost of shifting the BB at the origin to
future infinity, which does not look realistic (somewhere in the Universe the BB
would be going on forever).\footnote{\label{tube}What {\em looks} unrealistic at
first sight is not to be reflexively dismissed. This case does deserve a serious
investigation, but it will not be carried out in the present paper.} It still
needed to be investigated whether the ERS and BB detach if the hump on the BB
set is centered around a maximum or minimum of the areal radius that does not
coincide with the origin. In the present paper it is demonstrated by explicit
examples that this indeed happens. With the ERS and BB detached, each
blueshifted ray is building up its blueshift in a longer segment of its path.
Consequently, achieving the frequency range of the GRBs requires a lower or
narrower hump, and the angular size of the radiation source becomes much
smaller; see below. (The problem of too-long-lasting afterglows still remains
and is not discussed here.)

Sections \ref{QSSS} and \ref{symmetric} present the QSS model used in this
paper, the null geodesic equations and properties of redshift along them. In
Sec. \ref{nearmaxphi}, the parameters of the QSS region around a local minimum
of the areal radius are specified. In Sec. \ref{initpar}, a set of numerical
values of the parameters of the QSS region is chosen as a starting point for
improvements. In Sec. \ref{ERS}, the equation defining the ERS is derived and it
is shown that is has a unique solution at every $r$. In Sec. \ref{numex},
examples are given of QSS regions that generate sufficiently strong blueshift to
reach the frequency range of the GRBs. In Sec. \ref{decheight}, one of the
examples is further perfected to make the hump on the BB as low as possible.
With the ``best'' parameters, the BB hump has the diameter smaller than 1/5 and
the height smaller than 1/23 of that from Refs. \cite{Kras2018a,Kras2018b}. Sec.
\ref{backfromnow} investigates the numerical discrepancies between a ray
calculated from the central world line of the QSS region to the present time and
the same ray calculated backward from the present time to the starting point. It
is shown how the discrepancies can be minimised. In Sec. \ref{angrad}, the
angular radius of the gamma-ray source as seen by the present observer is
calculated -- it is $\approx 0.176^{\circ}$, and the whole sky could accommodate
more than 330,000 such sources without overlaps. This fits well with the
localisation errors for the 186 GRBs observed by the Large Area Telescope (LAT)
from 2008 to 2018 \cite{Ajel2019}, which are mostly contained between
0.04$^{\circ}$ and 1.0$^{\circ}$ (only 18 are greater). Sec. \ref{sumcon}
summarises the results of this paper and prospects for improving the model.

\section{The quasispherical Szekeres (QSS) spacetimes}\label{QSSS}

The signature is $(+, -, -, -)$, the coordinates are $\left(x^0, x^1, x^2,
x^3\right) = (t, r, x, y)$ or $(t, r, \vartheta, \varphi)$ and we assume the
cosmological constant $\Lambda = 0$.

The metric of the QSS spacetimes is \cite{Szek1975,Szek1975b,PlKr2006,Hell1996}
\begin{equation}\label{2.1}
{\rm d} s^2 = {\rm d} t^2 - \frac {\left(\Phi,_r - \Phi {\cal E},_r/{\cal
E}\right)^2} {1 + 2 E(r)} {\rm d} r^2 - \left(\frac {\Phi} {\cal E}\right)^2
\left({\rm d} x^2 + {\rm d} y^2\right),\ \ \ \ \
\end{equation}
\begin{equation}\label{2.2}
\rm{where} \qquad {\cal E} \df \frac S 2 \left[\left(\frac {x - P} S\right)^2 +
\left(\frac {y - Q} S\right)^2 + 1\right],
\end{equation}
$P(r)$, $Q(r)$, $S(r)$ and $E(r)$ being arbitrary functions such that $S \neq 0$
and $E \geq -1/2$ at all $r$.

The source in the Einstein equations is dust ($p = 0$) with the velocity field
$u^{\alpha} = {\delta_0}^{\alpha}$. The surfaces of constant $t$ and $r$ are
nonconcentric spheres, and $(x, y)$ are the stereographic coordinates on each
sphere. At a fixed $r$, they are related to the spherical coordinates by
\begin{eqnarray}\label{2.3}
x &=& P + S \cot(\vartheta/2) \cos \varphi, \nonumber \\
y &=& Q + S \cot(\vartheta/2) \sin \varphi.
\end{eqnarray}
The functions $(P, Q, S)$ determine the centres of the spheres in the spaces of
constant $t$ \cite{Kras2016b,BuSc2019}. Because of the non-concentricity, the
QSS spacetimes in general have no symmetry \cite{BoST1977}. The function
$\Phi(t,r)$ obeys
\begin{equation}\label{2.4}
{\Phi,_t}^2 = 2 E(r) + \frac {2 M(r)} {\Phi},
\end{equation}
where $M(r)$ is an arbitrary function. We will consider only models with $E >
0$, then the solution of (\ref{2.4}) is
\begin{eqnarray}\label{2.5}
\Phi(t,r) &=& \frac M {2E} (\cosh \eta - 1), \nonumber \\
\sinh \eta - \eta &=& \frac {(2E)^{3/2}} M \left[t - t_B(r)\right],
\end{eqnarray}
where $t_B(r)$ is an arbitrary function; $t = t_B(r)$ is the time of the BB
singularity, at which $\Phi(t_B, r) = 0$. We assume $\Phi,_t > 0$ (the Universe
is expanding).

The mass density implied by (\ref{2.1}) is

\begin{equation}\label{2.6}
\kappa \rho = \frac {2 \left(M,_r - 3 M {\cal E},_r / {\cal E}\right)} {\Phi^2
\left(\Phi,_r - \Phi {\cal E},_r / {\cal E}\right)}, \quad \kappa \df \frac {8
\pi G} {c^2}.
\end{equation}
This is a mass-dipole superposed on a spherical monopole \cite{DeSo1985},
\cite{Szek1975b}. The dipole vanishes where ${\cal E},_r = 0$. The density is
minimum where ${\cal E},_r/{\cal E}$ is maximum and vice versa \cite{HeKr2002}.

The arbitrary functions must be such that no shell-crossing singularities exist.
This is ensured by \cite{HeKr2002}:
\begin{eqnarray}
\frac {M,_r} {3M} &\geq& \frac {\cal P} S, \qquad \frac {E,_r} {2E} > \frac
{\cal P} S ~~~~\forall~r, \label{2.7} \\
{\rm where}\ \ {\cal P} &\df& {\sqrt{(S,_r)^2 + (P,_r)^2 + (Q,_r)^2}}.
\label{2.8}
\end{eqnarray}
These inequalities imply \cite{HeKr2002}
\begin{equation}\label{2.9}
\frac {M,_r} {3M} \geq \frac {{\cal E},_r} {\cal E}, \qquad \frac {E,_r} {2E} >
\frac {\Phi,_r} {\Phi} \qquad \forall~r.
\end{equation}

The extrema of ${\cal E},_r/{\cal E}$ with respect to $(x, y)$ are
\cite{HeKr2002}
\begin{equation}\label{2.10}
\left.\frac {{\cal E},_r} {\cal E}\right|_{\rm ex} = \varepsilon_2 \frac {\cal
P} S, \qquad \varepsilon_2 = \pm 1,
\end{equation}
with $+$ at a maximum and $-$ at minimum; they occur at
\begin{equation}\label{2.11}
x = P - \frac {\varepsilon_2 SP,_r} {{\cal P} + \varepsilon_2 S}, \qquad y = Q -
\frac {\varepsilon_2 SQ,_r} {{\cal P} + \varepsilon_2 S}.
\end{equation}

The model (\ref{2.1}) -- (\ref{2.2}) becomes axially symmetric when $P$ and $Q$
are constant. Then, $x$ and $y$ can be chosen such that $P = Q = 0$, and the set
$x = y = 0$ is the axis of symmetry. This is the case discussed here; then
\begin{equation}\label{2.12}
{\cal E} = \frac 1 {2 S}\ \left(x^2 + y^2 + S^2\right).
\end{equation}
In the axially symmetric case the maximally blueshifted rays stay in a fixed
hypersurface (they intersect the symmetry axis in every space of constant time
\cite{Kras2016b,Kras2018a}), which takes away one source of numerical errors.
Without any symmetry, since the direction of strongest blueshift is unstable
\cite{Kras2016b}, tracing the rays would require extreme numerical precision.
The form of the function $S$ is defined in Sec. \ref{nearmaxphi}, and conditions
(\ref{2.7}) are discussed there, too.

The following equation will be useful further on \cite{PlKr2006}:
\begin{equation}
\Phi,_{tr} = \frac {E,_r} {2E} \Phi,_t - \frac M {\Phi^2} \left[\left(\frac 3 2
\frac {E,_r} E - \frac {M,_r} M\right) \left(t - t_B\right) - t_{B,r}\right].
\label{2.13}
\end{equation}

The values of various parameters of the real Universe expressed in standard
physical units are too large numbers for numerical calculations. Therefore, the
numerical length unit (NLU) and the numerical time unit (NTU) were introduced in
Ref. \cite{Kras2014b}:
\begin{equation}\label{2.14}
1\ {\rm NTU} = 1\ {\rm NLU} = 3 \times 10^4\ {\rm Mpc} = 9.26 \times 10^{23}\
{\rm km} = 9.8 \times 10^{10}\ {\rm y}.
\end{equation}
The quantity $\kappa \rho$ in (\ref{2.6}) has the dimension of (length)$^{-2}$,
and in the units of (\ref{2.14}) its value at last scattering is
\cite{Kras2016a}
\begin{equation}
\kappa \rho_{\rm LS} = 56.1294161975316 \times 10^9 \ ({\rm NLU})^{-2}.
\label{2.15}
\end{equation}
In numerical calculations of past-directed null geodesics, $\kappa \rho$ is
calculated along. Where its value reaches (\ref{2.15}), that point is taken to
lie on the LSH, as explained in Sec. \ref{intro}, and the calculation stops.

The L--T models are the limit of (\ref{2.1}) -- (\ref{2.2}) at constant $(P, Q,
S)$. The Friedmann limit is obtained from QSS when $E / M^{2/3}$ and $t_B$ are
constant (then $(P, Q, S)$ can be made constant by a coordinate transformation).
QSS and Friedmann spacetimes can be matched at any constant $r$.

The spacetime model used further in this paper consists of a QSS region of
finite spatial volume matched to a Friedmann region across a $r = r_b =$
constant hypersurface. The metric in the Friedmann region is
\begin{equation}\label{2.16}
{\rm d} s^2 = {\rm d} t^2 - {\mathcal {R}}^2(t) \left[ \frac { {\rm d} r^2} {1 -
k r^2} + r^2 \left({\rm d} \vartheta^2 + \sin^2 \vartheta {\rm d}
\varphi^2\right)\right],\ \ \ \ \
\end{equation}
where the value of $k$ will be given in Sec. \ref{nearmaxphi}.

\section{Null geodesics in the axially symmetric QSS
spacetimes}\label{symmetric}

In (\ref{2.1}) -- (\ref{2.2}) $x = \infty$ and $y = \infty$ occur at the pole of
the stereographic projection. This is a coordinate singularity where
numerical integration of geodesics breaks down. So, we introduce the coordinates
$(\vartheta, \varphi)$ by
\begin{equation}\label{3.1}
x = S_b \cot(\vartheta/2) \cos \varphi, \qquad y = S_b \cot(\vartheta/2) \sin
\varphi,
\end{equation}
where $S_b$ is the value of $S$ at the Szekeres/Friedmann boundary
\begin{equation}\label{3.2}
S_b \df S(r_b)
\end{equation}
This changes (\ref{2.1}) and (\ref{2.2}) to
\begin{eqnarray}
{\rm d} s^2 &=& {\rm d} t^2 - \frac {{\cal N}^2 {\rm d} r^2} {1 + 2 E(r)} -
\left(\frac {\Phi} {\cal F}\right)^2 \left({\rm d} \vartheta^2 + \sin^2
\vartheta {\rm d} \varphi^2\right), \label{3.3}\\
{\cal F} &=& \frac {S_b} {2 S}\ (1 + \cos \vartheta) + \frac S {2 S_b}\ (1 -
\cos \vartheta), \label{3.4} \\
&{\rm where}& \qquad {\cal N} \df \Phi,_r - \Phi {\cal F},_r/{\cal F},
\label{3.5}
\end{eqnarray}
and the axis of symmetry is now at $\vartheta = \pi$ (where $x = y = 0$) and at
$\vartheta = 0$ (where both $x$ and $y$ become infinite -- in the stereographic
coordinates this is the antipodal point to $x = y = 0$).

In general, $(\vartheta, \varphi)$ are {\it not} the spherical coordinates
because ${\cal F}$ depends on $\vartheta$. The dipole equator ${\cal F},_r = 0$
is at $\cot (\vartheta_{\rm eq}/2) = S/S_b$. At $r = r_b$ ${\cal F} = 1$ and
$(\vartheta, \varphi)$ become the spherical coordinates with the origin at $r =
0$.

In the coordinates of (\ref{3.3}) -- (\ref{3.4}) equation (\ref{2.6}) becomes
\begin{equation}\label{3.6}
\kappa \rho = \frac {2 \left(M,_r - 3 M {\cal F},_r / {\cal F}\right)} {\Phi^2
\left(\Phi,_r - \Phi {\cal F},_r / {\cal F}\right)}.
\end{equation}

Along a geodesic, with $\lambda$ an affine parameter, we denote
\begin{equation}\label{3.7}
\left(k^t, k^r, k^{\vartheta}, k^{\varphi}\right) \df \dr {(t, r, \vartheta,
\varphi)} {\lambda}.
\end{equation}
Then, the geodesic equations for (\ref{3.3}) -- (\ref{3.4}) are
\begin{eqnarray}
\dr {k^t} {\lambda} &+& \frac {{\cal N} {\cal N},_t} {1 + 2E} \left(k^r
\right)^2 + \frac{\Phi {\Phi,_t}}{{\cal F}^2} \left[\left(k^{\vartheta}\right)^2
+ \sin^2 \vartheta \left(k^{\varphi}\right)^2\right] = 0, \label{3.8} \\
\dr {k^r} {\lambda} &+& 2 \frac {{\cal N},_t} {\cal N} k^t k^r + \left(\frac
{{\cal N},_r} {\cal N} - \frac {E,_r} {1 + 2E}\right) \left(k^r\right)^2 + 2
\frac {S,_r \sin \vartheta \Phi} {S {\cal F}^2 {\cal N}}\ k^r k^{\vartheta}
\nonumber \\
&-& \frac {\Phi (1 + 2E)} {{\cal F}^2 {\cal N}}
\left[\left(k^{\vartheta}\right)^2 + \sin^2 \vartheta
\left(k^{\varphi}\right)^2\right] = 0, \label{3.9} \\
\dr {k^{\vartheta}} {\lambda} &+& 2 \frac {\Phi,_t} {\Phi} k^t k^{\vartheta} -
\frac {S,_r \sin \vartheta {\cal N}} {S \Phi (1 + 2E)}\ \left(k^r\right)^2 + 2
\frac {\cal N} {\Phi} k^r k^{\vartheta} \nonumber \\
&+& \frac {{\cal F},_{\vartheta}} {\cal F}\ \left[- \left(k^{\vartheta}\right)^2
+ \sin^2 \vartheta \left(k^{\varphi}\right)^2\right] - \cos \vartheta \sin
\vartheta \left(k^{\varphi}\right)^2 = 0, \label{3.10} \\
\dr {k^{\varphi}} {\lambda} &+& 2 \frac {\Phi,_t} {\Phi} k^t k^{\varphi} + 2
\frac {\cal N} {\Phi} k^r k^{\varphi} + 2 \left[\frac {\cos \vartheta} {\sin
\vartheta}
 - \frac {{\cal F},_{\vartheta}} {\cal F}\right] k^{\vartheta} k^{\varphi} = 0.
 \label{3.11}
\end{eqnarray}
The geodesics determined by (\ref{3.8}) -- (\ref{3.11}) are null when
\begin{equation}\label{3.12}
\left(k^t\right)^2 - \frac {{\cal N}^2 \left(k^r\right)^2} {1 + 2E(r)} -
\left(\frac {\Phi} {\cal F}\right)^2 \left[\left(k^{\vartheta}\right)^2 + \sin^2
\vartheta \left(k^{\varphi}\right)^2\right] = 0.
\end{equation}

On past-directed rays $k^t < 0$, and $\lambda$ along each of them can be chosen
such that at the observation point
\begin{equation}\label{3.13}
k^t_o = -1.
\end{equation}
(On future-directed rays $k^t > 0$ and a convenient choice of $\lambda$ is
$k^t_e = +1$.)

In the Friedmann region we choose the coordinates so that \cite{Kras2016a}
\begin{equation}\label{3.14}
S = S_b.
\end{equation}
Then, throughout the Friedmann region, ${\cal F} = 1$ and $(\vartheta, \varphi)$
are the spherical coordinates. They coincide with the coordinates of the QSS
region at $r = r_b$.

To calculate $k^r$ on nonradial rays, (\ref{3.12}) will be used, which is
insensitive to the sign of $k^r$. This sign will be changed by the numerical
program integrating \{(\ref{3.8}), (\ref{3.10}) -- (\ref{3.12})\} at each point
where $k^r$ reaches zero.

Note that $\vartheta \equiv 0$ and $\vartheta \equiv \pi$ are solutions of
(\ref{3.10}). These axial rays intersect every space of constant $t$ on the
symmetry axis.

Along a ray emitted at $P_e$ and observed at $P_o$, with $k^{\alpha}$ being
affinely parametrised, we have
\begin{equation}\label{3.15}
1 + z = \frac {\left(u_{\alpha} k^{\alpha}\right)_e} {\left(u_{\alpha}
k^{\alpha}\right)_o},
\end{equation}
where $u_{\alpha}$ are four-velocities of the emitter and of the observer
\cite{Elli1971}. In our case, both the emitter and the observer comove with the
cosmic matter, so $u_{\alpha} = {\delta^0}_{\alpha}$, and the affine parameter
is chosen so that (\ref{3.13}) holds; then
\begin{equation}\label{3.16}
1 + z = - {k_e}^t.
\end{equation}
Equation (\ref{3.11}) has the first integral:
\begin{equation}\label{3.17}
k^{\varphi} \sin^2 \vartheta \Phi^2 / {\cal F}^2 = J_0,
\end{equation}
where $J_0$ is constant along each geodesic. Using (\ref{3.17}), eq.
(\ref{3.12}) implies
\begin{equation}\label{3.18}
(k^t)^2 = \frac {{\cal N}^2 \left(k^r\right)^2} {1 + 2E} + \left(\frac {\Phi}
{\cal F}\right)^2 \left(k^{\vartheta}\right)^2 + \left(\frac {J_0 {\cal F}}
{\sin \vartheta \Phi}\right)^2.
\end{equation}
At the observation/emission point, (\ref{3.13})/(\ref{3.16}), respectively,
apply. Equations (\ref{3.18}) and (\ref{3.16}) show that for rays emitted at the
BB, where $\Phi = 0$, the observed $z$ is infinite when $J_0 \neq 0$. A
necessary condition for infinite blueshift ($1 + z_o = 0$) is thus $J_0 = 0$, so

(a) either $k^{\varphi} = 0$,

(b) or $\vartheta = 0, \pi$ along the ray ((\ref{3.17}) implies
$J_0/\sin \vartheta \to 0$ when $\vartheta \to 0, \pi$).

\noindent Condition (b) appears to be also sufficient, but so far this has been
demonstrated only numerically in concrete examples of QSS models
(\cite{Kras2016b,Kras2018a}).

Condition (a) is {\em not} sufficient, and Ref. \cite{Kras2016b} contains
numerical counterexamples: there exist rays that proceed in a surface of
constant $\varphi$, but approach the BB with $z \to \infty$; the value of
$\vartheta$ along them changes and is different from $0, \pi$. For those rays,
(\ref{3.18}) with $J_0 = 0$ implies one more thing
\begin{eqnarray} \label{3.19}
&& {\rm If\ } \lim_{t \to t_B} z = \infty\ {\rm and}\ \lim_{t \to t_B}
\left|k^r\right| < \infty \nonumber \\
&& {\rm then}\ \lim_{t \to t_B} k^{\vartheta} = \pm \infty,
\end{eqnarray}
i.e., such rays approach the BB tangentially to the surfaces of constant $r$.

Consider a ray proceeding from event $P_1$ to $P_2$ and then from $P_2$ to
$P_3$. Let the redshifts acquired in the intervals $[P_1, P_2]$, $[P_2, P_3]$
and $[P_1, P_3] = [P_1, P_2] \cup [P_2, P_3]$ be $z_{12}$, $z_{23}$ and
$z_{13}$, respectively. Then, from (\ref{3.15}),
\begin{equation} \label{3.20}
1 + z_{13} = \left(1 + z_{12}\right) \left(1 + z_{23}\right).
\end{equation}
Thus, for a ray proceeding to the past from $P_1$ to $P_2$, and then back to the
future from $P_2$ to $P_1$:
\begin{equation}\label{3.21}
1 + z_{12} = \frac 1 {1 + z_{21}}.
\end{equation}

\section{Relations around a spatial minimum of $\Phi(t,r)$}\label{nearmaxphi}

For the metric (\ref{2.1}) -- (\ref{2.2}), in the orthonormal tetrad of
differential forms:
\begin{eqnarray}
{\rm e}^0 &=& {\rm d} t, \quad \quad {\rm e}^1 = \frac F {\sqrt{1 + 2E}}\ {\rm
d} r, \quad {\rm e}^2 = \frac {\Phi} {\cal E}\ {\rm d} x, \quad {\rm e}^3 =
\frac {\Phi} {\cal E}\ {\rm d} y, \label{4.1} \ \ \ \ \  \\
&{\rm where}& \ F \df \Phi,_r - \Phi {\cal E},_r / {\cal E}, \label{4.2}
\end{eqnarray}
the tetrad components of the curvature tensor are
\begin{eqnarray}
&& R_{0101} = \frac {2M} {\Phi^3} - \frac {M,_r - 3M {\cal E},_r / {\cal E}}
{\Phi^2 F}, \label{4.3} \\
&& R_{0202} = R_{0303} = \frac 1 2\ R_{2323} = - \frac M {\Phi^3}, \label{4.4}
\\
&& R_{1212} = R_{1313} = \frac M {\Phi^3} - \frac {M,_r - 3M {\cal E},_r / {\cal
E}} {\Phi^2 F}. \label{4.5}
\end{eqnarray}
These are scalars, so any scalar polynomial in curvature components is
determined by them.

The metric (\ref{2.1}) has a singularity where $F = 0$, but as seen from the
above, this will not be a curvature singularity if $M,_r - 3M {\cal E},_r /
{\cal E}$ has there a zero of the same order. Such a location is either a neck
(where $2E + 1 = 0$ of the same order) \cite{PlKr2006,HeKr2002} or a local
spatial extremum of $\Phi$. In those cases, $F = 0$ is just a coordinate
singularity.

For a neck to exist, $E$ must be negative in its neighbourhood. To consider this
case, we would have to either take a different background $E$ from those
considered in Refs. \cite{Kras2016a} -- \cite{Kras2018b} (where $E$ was positive
and Friedmannian) or allow the sign of $E$ to vary within the QSS region. In the
first case, we would give up on the correspondence with the previous papers, the
second case would introduce an additional complication. So, for this exploratory
investigation, we will consider a spatial extremum of $\Phi(t, r)$.

The equations  $F = 0$ and $M,_r - 3M {\cal E},_r / {\cal E} = 0$ can be
simultaneously fulfilled only if, at that location,
\begin{equation}\label{4.6}
M,_r = E,_r = P,_r = Q,_r = S,_r = {\dril {t_B} r} = 0,
\end{equation}
and then the extremum is comoving with the cosmic dust \cite{PlKr2006,HeKr2002}.
All zeros must be of the same order. If the extremum does not coincide with the
origin $\Phi = 0$, then $M$ at it must be nonzero -- see (\ref{2.5}).

The metric (\ref{2.1}) -- (\ref{2.2}) is covariant with transformations of the
form $r = f(r')$, where $f$ is an arbitrary function. Consequently, we can
choose $r$ such that the extremum is at $r = 0$. Suppose that all the zeros in
(\ref{4.6}) are of order $(n - 1)$, where $n \geq 2$ is a natural number to be
chosen later. The simplest $M$, $E$, $t_B$ and $S$ with this property have the
following form:
\begin{eqnarray}
M &=& M_{\rm ext} + D r^n, \label{4.7} \\
E &=& E_{\rm ext} + A r^n, \label{4.8} \\
t_B &=& t_{\rm Bext} - B r^n, \label{4.9} \\
S &=& \sqrt{r^n + a^n}, \label{4.10}
\end{eqnarray}
where the subscript ``{\it e}'' stands for ``at extremum of $\Phi$'', and all
the symbols newly introduced here are constants. The signs in (\ref{4.7}) --
(\ref{4.9}) were chosen such that $D$, $A$ and $B$ are all positive for a
spatial minimum of $\Phi$ at $r = 0$. Also, $M_{\rm ext} = M(0) > 0$ (because $M
> 0$ always) and $E_{\rm ext} = E(0) > 0$ because we now follow the $E > 0$ model. The
form of $S$ was chosen for correspondence with Refs.
\cite{Kras2016b,Kras2018a,Kras2018b} when $n = 2$. We shall consider a minimum
because this leads to simpler formulae (a maximum is left for a later paper, if
anybody cares to write it).

For a spatial minimum of $\Phi$, a neighbourhood of $r = 0$ exists in which, at
a fixed $t = t_o$, $\Phi,_r > 0$. Then, to avoid shell crossings in this
neighbourhood, the following conditions must be obeyed (\cite{HeKr2002} with $P
= Q = 0$):
\begin{eqnarray}
&& M,_r > 0, \qquad E,_r > 0, \qquad t_{B,r} < 0,\ \ \ \ \
\label{4.11} \\
&& \frac {S,_r} S < \frac {M,_r} {3M}, \qquad \frac {S,_r} S < \frac {E,_r}
{2E}. \label{4.12}
\end{eqnarray}
Since $M > 0$ and we assume $E > 0$ (for correspondence with earlier papers),
the equations above imply
\begin{eqnarray}
&& D > 0, \qquad A > 0, \qquad B > 0, \label{4.13} \\
&& M_{\rm ext} < \tfrac 1 3\ D \left(2 a^n - r^n\right), \qquad E_{\rm ext} <
Aa^n. \label{4.14}
\end{eqnarray}
Somewhere in the range of $r$ determined by (\ref{4.14}), the QSS region will be
matched to a Friedmann background, where $E(r) = - \tfrac 1 2\ kr^2$ and $M(r) =
M_0 r^3$, with constant $k$ and $M_0$. Let the matching hypersurface be $r =
r_b$. Since $r \leq r_b$ in the QSS region, a sufficient condition for the first
of (\ref{4.14}) is
\begin{equation}\label{4.15}
M_{\rm ext} < \tfrac 1 3\ D \left(2 a^n - {r_b}^n\right).
\end{equation}

The value of $k$ is in principle arbitrary, but, for correspondence (we wish to
have the same Friedmann background as in Refs. \cite{Kras2016a} --
\cite{Kras2018b}), we choose
\begin{equation}\label{4.16}
k = -0.4.
\end{equation}
Also for correspondence, we choose\footnote{$M = Gm/c^2$, where $m$ is mass, so
$M$ is measured in length units. Since $r$ is dimensionless, the units of $M$,
$D$ and $M_{\rm ext}$ are also NLU.}
\begin{equation}\label{4.17}
M_0 = 1\ {\rm NLU}.
\end{equation}
At the QSS/Friedmann boundary we must thus have
\begin{eqnarray}
E_{\rm ext} + A {r_b}^n &=& - \tfrac 1 2\ k {r_b}^2, \label{4.18} \\
M_{\rm ext} + D {r_b}^n &=& M_0 {r_b}^3. \label{4.19}
\end{eqnarray}
The $M_{\rm ext}$ and $D$ must be chosen in agreement with (\ref{4.15}) and
(\ref{4.19}), and for $E_{\rm ext}$ consistency between (\ref{4.14}) and
(\ref{4.18}) imposes the condition
\begin{equation}\label{4.20}
E_{\rm ext} = - \tfrac 1 2\ k {r_b}^2 - A {r_b}^n < Aa^n,
\end{equation}
which is equivalent to
\begin{equation}\label{4.21}
A > \frac {- \frac 1 2\ k {r_b}^2} {{r_b}^n + a^n} \df {\overline A}.
\end{equation}

\section{The initial choice of parameter values}\label{initpar}

As a test of the model, the numerical calculation of blueshift on the rays
emitted at the spatial minimum of $\Phi(t,r)$ was at first done with the values
of the parameters in the QSS region that were not too different from those in
the previous papers \cite{Kras2016a,Kras2018a}. The QSS/Friedmann boundary is
here at $r = r_b$, and in Ref. \cite{Kras2018a} it was at
\begin{equation}\label{5.1}
r = B_1 + A_1 = 0.015 + 10^{-10},
\end{equation}
so a realistic first choice is
\begin{equation}\label{5.2}
r_b = 0.015.
\end{equation}
The BB time at $r = 0$ is here $t_{\rm Bext}$, at $r = r_b$ it is $t_{\rm Bext}
- B {r_b}^n$. The difference, $B {r_b}^n$, is the height of the hump on the BB.
In Ref. \cite{Kras2018a}, the height was
\begin{equation}\label{5.3}
B_0 + A_0 = 0.000126\ {\rm NTU} \df H.
\end{equation}
So, we impose the condition
\begin{equation}\label{5.4}
B {r_b}^n = H.
\end{equation}

All these conditions now have to be made into a self-consistent set. So, the
initial prescription for constructing a QSS region free of shell crossings
around a spatial minimum of $\Phi$ is:

($i$) Select $n$. We choose $n = 6$, since in previous papers the BB profile was
a curve of degree 6.

($ii$) Choose $r_b = 0.015$, as in (\ref{5.2}).

($iii$) With $H$ given by (\ref{5.3}), $B$ is\footnote{The values of $B$,
$E_{\rm ext}$ and $M_{\rm ext}$ in (\ref{5.5}), (\ref{5.8}) and (\ref{5.10})
were calculated in the Fortran program at double precision.}
\begin{equation}\label{5.5}
B = H / {r_b}^n = 11\ 061\ 728.395061729\ {\rm NTU}.
\end{equation}

% $(0.015)^6 = 0,000000000011390625$
% $1/(0.015)^6 = 87 791 495 198,902606310013717421125$
% calculated by WinEdt $B = 11061728,395061728395061728395062$

($iv$) Choose $a^n$. We take it the same as $a^2$ in the previous papers:
\begin{equation}\label{5.6}
a^n = 0.001.
\end{equation}

($v$) Choose $A > {\overline A}$ in agreement with (\ref{4.21}).\footnote{The
value of ${\overline A}$ found by the calculator of the WinEdt program
\cite{Winedt} is $0.04499999948742188083858513857299$. This calculator is {\em
more precise} than Fortran. } We choose
\begin{equation}\label{5.7}
A = 0.05.
\end{equation}
With $k = -0.4$, $n = 6$ and $r_b \leq 0.015$, the ${\overline A}$ defined in
(\ref{4.21}) has $\dril {\overline A} {r_b} > 0$. Consequently, when $r_b$ is
made smaller, $A = 0.05$ will fulfil (\ref{4.21}) with a wider margin. With $k$,
$r_b$ and $A$ already defined, we obtain from (\ref{4.20})
\begin{equation}\label{5.8}
E_{\rm ext} = 4.4999999430468751 \times 10^{-5}.
\end{equation}
% $E_e = 0,000044999999430468751$
% $A a^n = 0,00005$, so indeed $E_e < A a^n$
This obeys $E_{\rm ext} < A a^n$ since $A a^n = 5 \times 10^{-5}$. Similarly to
what happens with ${\overline A}$, with the values of $n$, $k$ and $A$ given
above, $E_{\rm ext}$ is an increasing function of $r_b$. Consequently,
(\ref{4.20}) will be fulfilled with $r_b < 0.015$.

($vi$) Choose $D > 0$. There is no other condition on $D$, so we take
\begin{equation}\label{5.9}
D = 1\ {\rm NLU}.
\end{equation}

($vii$) Now $M_{\rm ext}$ is determined by (\ref{4.19}). With the already-chosen
values of $M_0$, $r_b$ and $D$ we have
\begin{equation}\label{5.10}
M_{\rm ext} = 3.3749886093750001 \times 10^{-6}\ {\rm NLU},
\end{equation}
while $M_u \df \tfrac 1 3 D \left(2 a^n - {r_b}^n\right) =
0.0006666666628697916...$, so (\ref{4.15}) is obeyed. Also, $\dril {M_{\rm ext}}
{r_b} > 0$ while $M_u$ becomes greater when $r_b$ decreases, so with $r_b <
0.015$ $M_{\rm ext}$ will be smaller than (\ref{5.10}) and will continue to obey
(\ref{4.15}).

As in \cite{Kras2016a} -- \cite{Kras2018b}, for the BB time in the Friedmann
background we take
\begin{equation}\label{5.11}
t_{\rm Bf} = -0.13945554689046649\ {\rm NTU} \approx -13.67 \times 10^9\ {\rm
years};
\end{equation}
see Ref. \cite{Kras2018b} for justification. So,
\begin{equation}\label{5.12}
t_{\rm Bext} = t_{\rm Bf} + H = -0.13932954689046649\ {\rm NTU}.
\end{equation}

Caution must be exercised while calculating $k^r$ from (\ref{3.12}). If $r = 0$
is not a neck, then, with the $r$-coordinate used so far, ${\cal N}|_{r = 0} =
0$, but $1 + 2E|_{r = 0} \neq 0$ and $\left.k^r\right|_{r = 0}$ comes out
infinite. Therefore, in using this equation, one must change the $r$-coordinate
to $\overline{r} = r^n$, at least in a neighbourhood of $r = 0$. Thus, the order
of zero of the derivatives in (\ref{4.6}) is in fact irrelevant: one can do the
transformation $\overline{r} = r^n$, and then $r' = {\overline{r}}^{1/m}$ with
any $m \neq n$ -- the resulting $\Phi,_{r'}$ will have a zero at $r' = 0$ of
order $m \neq n$, but the metric will be just a coordinate transform of the
original one. However, with a changed $n$ the values of the other parameters of
the QSS region will be also changed.

\section{The Extremum Redshift Surface}\label{ERS}

Consider a null geodesic that stays in one of the two surfaces:
\begin{equation}\label{6.1}
\{\cos \vartheta, \varphi\} = \{- \varepsilon, {\rm constant}\},
\end{equation}
where $\varepsilon = \pm 1$, i.e., $\vartheta = \pi$ or $\vartheta = 0$,
respectively. Such geodesics obey (\ref{3.10}) and (\ref{3.11}) provided the
limit $\varphi =$ constant in (\ref{3.11}) is taken first. Along the direction
$\vartheta = \pi$ ($\varepsilon = +1$) the dipole is maximum, along the other
one ($\vartheta = 0, \varepsilon = -1$) it is minimum.

All along such a geodesic, $k^r \neq 0$ because wherever $k^r = 0$ the geodesic
would be timelike, so $r$ can be used as a parameter. Assume the geodesic is
past-directed so that (\ref{3.16}) applies. Then we obtain from (\ref{3.8})
using (\ref{3.16})
\begin{equation}\label{6.2}
\dr z r = \frac {{\cal N} {\cal N},_t} {1 + 2E}\ k^r.
\end{equation}
Since ${\cal N} \neq 0$ from no-shell-crossing conditions \cite{HeKr2002} and
$k^r \neq 0$, the extrema of $z$ on such a geodesic occur where
\begin{equation}\label{6.3}
{\cal N},_t \equiv \Phi,_{tr} - \Phi,_t {\cal F},_r/{\cal F} = 0.
\end{equation}
In deriving (\ref{6.3}), the constant $\varphi$ was arbitrary. Thus, the set
defined by (\ref{6.3}) is 2-dimensional; it is the Extremum Redshift
Surface (ERS) \cite{Kras2016b}.

With (\ref{6.1}) obeyed, ${\cal F},_r/{\cal F} = \varepsilon S,_r / S$. Using
(\ref{2.13}), Eq. (\ref{6.3}) becomes
\begin{equation}\label{6.4}
\left(\frac {E,_r} {2E} - \varepsilon \frac {S,_r} S\right) \Phi,_t - \frac M
{\Phi^2} \left[\left(\frac 3 2 \frac {E,_r} E - \frac {M,_r} M\right) \left(t -
t_B\right) - t_{B,r}\right] = 0.
\end{equation}
Substituting for $\Phi$, $\Phi,_t$ and $(t - t_B)$ from (\ref{2.5}), Eq.
(\ref{6.4}) is transformed to
\begin{eqnarray}
&& \sqrt{2E} \left[\left(\frac {E,_r} {2E} - \varepsilon \frac {S,_r} S\right)
\sinh \eta \cosh \eta + \left(- 2 \frac {E,_r} E + \frac {M,_r} M + \varepsilon
\frac {S,_r} S\right) \sinh \eta\right. \nonumber\\
&& + \left.\left(\frac 3 2 \frac {E,_r} E - \frac {M,_r} M\right) \eta\right] +
\frac {(2E)^2} M\ t_{B,r} = 0.\ \ \  \label{6.5}
\end{eqnarray}
This is the equation of the ERS. In the limit $S,_r = 0$ it reproduces the
equation of the Extremum Redshift {\em Hyper}surface (ERH) of Ref.
\cite{Kras2014d}.

Equation (\ref{6.5}) implies that, with $S(r)$ given by (\ref{4.10}), the ERS
coincides with the BB {\it at the origin}\footnote{The origin is not to be
confused with the BB extremum considered further on.} $r = r_{\rm or}$ if and
only if $\lim_{r \to r_{\rm or}} \left[(r - r_{\rm or}) \dril {t_B} r\right] =
0$; see Appendix \ref{ERSatorig}. Consequently, the two sets are ``unglued'' at
$r = r_{\rm or}$ if and only if $\lim_{r \to r_{\rm or}} \left[(r - r_{\rm or})
\dril {t_B} r\right] = C \neq 0$. Then, in a neighbourhood of the origin, the
function $t_B(r)$ behaves like [$- \ln (r - r_{\rm or})$], so $\lim_{r \to
r_{\rm or}} t_B(r) = \infty$. This means that somewhere in the Universe the BB
would be still going on now (and would go on forever). Whether this is
``plausible'' or not, such a geometry deserves to be
investigated, see footnote \ref{tube}.

Substituting (\ref{4.7}) -- (\ref{4.10}) in (\ref{6.5}) and canceling $n r^{n -
1}$ we obtain
\begin{equation}
{\cal H}(r, \eta) = F_4(r), \label{6.6}
\end{equation}
where
\begin{eqnarray}
{\cal H}(r, \eta) &=& F_1(r) \sinh \eta \cosh \eta + F_2(r) \sinh \eta + F_3(r)
\eta, \label{6.7} \\
F_1(r) &=& \frac A {E_{\rm ext} + Ar^n} - \frac {\varepsilon} {r^n + a^n} \equiv
\frac {(1 - \varepsilon) A r^n + Aa^n - \varepsilon E_{\rm ext}} {\left(E_{\rm
ext} + Ar^n\right) \left(r^n + a^n\right)} , \label{6.8} \\
F_2(r) &=& - \frac {4A} {E_{\rm ext} + Ar^n} + \frac {2D} {M_{\rm ext} + Dr^n} +
\frac {\varepsilon} {r^n + a^n}\ ,\ \ \ \  \label{6.9} \\
F_3(r) &=& \frac {3A} {E_{\rm ext} + Ar^n} - \frac {2D} {M_{\rm ext} + Dr^n}\ ,
\label{6.10} \\
F_4(r) &=& \frac {2^{5/2} \left(E_{\rm ext} + Ar^n\right)^{3/2} B} {M_{\rm ext}
+ Dr^n}. \label{6.11}
\end{eqnarray}
Taking (\ref{6.6}) at $r = 0$ we see that $\eta = 0$ fulfils it only when
$E_{\rm ext} B = 0$ -- only then the ERS coincides with the BB at the BB
extremum. If we wish to unglue these two sets at that point, we must take $B
E_{\rm ext} \neq 0$ in (\ref{4.8}) -- (\ref{4.9}). Our choice (\ref{5.5}) and
(\ref{5.8}) guarantees this.

Extrema of redshift exist also along other directions than $\vartheta = 0$ and
$\vartheta = \pi$, as was demonstrated by numerical examples in Refs.
\cite{Kras2016a} -- \cite{Kras2018b}, but a general equation defining their loci
remains to be derived.

With the values of the parameters in (\ref{5.2}) -- (\ref{5.12}), one can verify
that ${\cal H} > 0$ and $\pdril {\cal H} {\eta} > 0$ for all $\eta
> 0$, see Appendix \ref{solvable}. Since $F_4(r) > 0$ for all $r > 0$ and is
independent of $\eta$, the following is true: at $\eta = 0$, ${\cal H} = 0 <
F_4(r)$ for all $r > 0$, at $\eta \to \infty$, ${\cal H} \to +\infty$, so ${\cal
H} > F_4(r)$ at all finite $r > 0$ for sufficiently great $\eta$. Thus,
somewhere in the range $\eta \in (0, \infty)$ Eq. (\ref{6.6}) has a unique
solution for $\eta$ at any finite $r > 0$. The initial $\eta > 0$ for the
numerical program solving (\ref{6.6}) is found also in Appendix \ref{solvable}.

\section{The numerical values of blueshift}\label{numex}

The formulae in Secs. \ref{nearmaxphi} and \ref{initpar} presented those
features of the QSS region that will not vary between numerical experiments.
This section presents the first numerical implementation and its consecutive
improvements. The aim of the whole action is to (1) achieve the lowest possible
value of $1 + z$ with a given set of parameters by fine-tuning the point where
the ray intersects the $r = 0$ line, and then (2) decrease the diameter and
height of the BB hump as much as possible while keeping $1 + z$ in the range
\cite{Kras2016a}
\begin{equation}\label{7.1}
2.56 \times 10^{-8} < 1 + z < 1.689 \times 10^{-5}
\end{equation}
needed to blueshift the emission frequencies of hydrogen and helium atoms (the
dominating matter components in the epoch of last scattering) to the range of
frequencies of the observed GRBs. The lower end of this range corresponds to the
highest-frequency emission radiation being blueshifted to the highest energy of
the observed GRBs, the upper end of (\ref{7.1}) corresponds to matching the
lowest-frequency ends of the two bands. The aim of the current paper is to find
out how low and thin the BB hump can be made while (\ref{7.1}) still holds.

\subsection{Model 1}\label{firstnum}

With the numerical values of the parameters given in Secs. \ref{nearmaxphi} and
\ref{initpar}, a light ray running in the surface (\ref{6.1}) with $\varepsilon
= +1$, sent to the past from $r = 0$ at $t = t_B(0) + \Delta t_{c1}$, where
\begin{equation}\label{7.2}
\Delta t_{c1} = 0.00000449960000\ {\rm NTU},
\end{equation}
crossed the LSH with
\begin{equation}\label{7.3}
1 + z_{p1} = 8.1259273421174782 \times 10^{-8}
\end{equation}
relative to the initial point. Achieving a still smaller $1 + z$ was probably
possible, but would require extreme numerical precision to correctly catch the
$(t, r)$ point where the ray intersects the LSH (this is because the ray and the
LSH intersect at a very small angle, see Fig. \ref{down3}). The result
(\ref{7.3}) was comparable to the best one achieved in Ref. \cite{Kras2018a} and
was good enough as a starting point for improvements of the BB profile.

A ray sent from the same initial point to the future, in the surface
$\{\vartheta, \varphi\} = \{0, {\rm constant}\}$, reached the present
time\footnote{Because of numerical inaccuracies, the ray overshot the present
time $t = 0$ by $t_{\rm now1}$ given by (\ref{7.4}), and the other numbers in
(\ref{7.4}) -- (\ref{7.5}) refer to that endpoint.} with
\begin{eqnarray}\label{7.4}
1 + z_{f1} &=& 55.299746938015609, \nonumber \\
t_{\rm now1} &=& 5.0391335364848865 \times 10^{-11}\ {\rm NTU}, \nonumber \\
r_{\rm now1} &=& 0.89044002852488546.
\end{eqnarray}
In the following, the concatenation of the two rays described above
will be called Ray 1. On it, the blueshift between the LSH and $t_{\rm now1}$
was
\begin{equation}\label{7.5}
1 + z_{t1} = \left(1 + z_{f1}\right) \times \left(1 + z_{p1}\right) =
4.49361725656 \times 10^{-6}.
\end{equation}
This is $\approx 0.2855$ of the value obtained with a BB hump of the same height
and nearly the same diameter but centered around the origin (Eq. (8.12) in Ref.
\cite{Kras2018a}). Thus, a BB hump around a spatial minimum of $\Phi$ generates
blueshifts more efficiently than a similar hump around the origin. The reason
for this is explained at the end of the present section.

%The exact values were
%upward redshift 0.018083265392170422
%downward redshift 1/0.018083265392170422 = 55,299746938015609
%1 + z_t = \left(1 + z_f\right) \times \left(1 + z_p\right) =
%8,1259273421174782 \times 10^{-8} \times 55,299746938015609
%449,36172565579833112661486331722 \times 10^{-8}

%The exact value of the fraction is 0,28548919949174078780177890724269

\subsection{Model 2}\label{secondnum}

In the second numerical experiment, the radius of the BB hump was decreased to
$r_{b2} = 0.01$, which changed the values of $M_{\rm ext}$, $E_{\rm ext}$ and
$B$ to
\begin{eqnarray*}
{\rm second}\ M_{\rm ext} &=& 10^{-6} - 10^{-12}, \\
{\rm second}\ E_{\rm ext} &=& 0.00001999999995, \\
{\rm second}\ B &=& 1.26 \times 10^8.
\end{eqnarray*}
The other parameters did not change, and, as predicted in Sec. \ref{initpar},
the inequalities (\ref{4.15}) and (\ref{4.20}) still held. On a ray sent to the
past from $r = 0$ in the direction of dipole maximum ((\ref{6.1}) with
$\vartheta = \pi$) the parameter $\Delta t_{c2}$ that resulted in the smallest
$1 + z$ at the LSH was
\begin{equation}\label{7.6}
\Delta t_{c2} = 0.00000133331600\ {\rm NTU},
\end{equation}
and the smallest $1 + z$ was
\begin{equation}\label{7.7}
1 + z_{p2} = 7.5237815977402533 \times 10^{-11}.
\end{equation}
The ray sent to the future from the same initial point in the direction of the
dipole minimum ($\vartheta = 0$) overshot the present time by $t_{\rm now2}$
given below. The parameters of the endpoint were
\begin{eqnarray}
1 + z_{f2} &=& 56.981145007279054, \label{7.8} \\
t_{\rm now2} &=& 4.3253430781086085 \times 10^{-10}\ {\rm NTU}, \label{7.9} \\
r_{\rm now2} &=& 0.88867576379669344. \label{7.10}
\end{eqnarray}
The total blueshift between the LSH and $t_{\rm now2}$ was thus
\begin{equation}\label{7.11}
1 + z_{t2} = 4.2871369 \times 10^{-9}.
\end{equation}
In the following, the concatenation of these two rays will be called Ray 2.

%B parameter on Ray 2 = $1.26 \times 10^8$
%M_e parameter on Ray 2 = $0.000000999999$

%56,981145007279054 \times 7,5237815977402533 \times 10^{-11} =
%428,71369022393505815818608174438 \times 10^{-11}

\subsection{Model 3}\label{thirdnum}

In the third numerical experiment, $r_b$ was decreased to $r_{b3} = 0.005$. The
new values of $M_{\rm ext}$, $E_{\rm ext}$ and $B$ became
\begin{eqnarray*}
{\rm third}\ M_{\rm ext} &=& 1.24999984375 \times 10^{-7}, \\
{\rm third}\ E_{\rm ext} &=& 9.9999999921875 \times 10^{-7}, \\
{\rm third}\ B &=& 8.064 \times 10^9,
\end{eqnarray*}
%(5 x 10^{-3})^3 = 0,000000125
%(5 x 10^{-3})^6 = 0,000000000000015625
% M_e = (5 x 10^{-3})^3 - (5 x 10^{-3})^6 = 0,000000124999984375
%(5 x 10^{-3})^2 = 0,000005
%0,05 x (5 x 10^{-3})^6 = 0,00000000000000078125
%E_e = 0,2 x (5 x 10^{-3})^2 - 0,05 x (5 x 10^{-3})^6
%= 0,000001 - 0,05 x (5 x 10^{-3})^6 = 0,00000099999999921875
which again preserved (\ref{4.15}) and (\ref{4.20}). The past-directed ray sent
from $r = 0$ in the surface (\ref{6.1}) along $\vartheta = \pi$ had the smallest
$1 + z$ at the LSH when
\begin{equation}\label{7.12}
\Delta t_{c3} = 0.00000016666400\ {\rm NTU};
\end{equation}
and the blueshift on it at the LSH was
\begin{equation}\label{7.13}
1 + z_{p3} = 1.8781317501215256 \times 10^{-8}.
\end{equation}
The ray sent to the future from the same initial point in the direction of the
dipole minimum $(\vartheta = 0)$ overshot the present time by $t_{\rm now3}$,
with
\begin{eqnarray}
1 + z_{f3} &=& 73.679048074068589, \label{7.14} \\
t_{\rm now3} &=& 5.0921478176623031 \times 10^{-10}\ {\rm NTU},\ \ \ \
\label{7.15} \\
r_{\rm now3} &=& 0.88725616206450841. \label{7.16}
\end{eqnarray}
The total $z$ between the LSH and $t_{\rm now3}$ was thus
\begin{equation}\label{7.17}
1 + z_{t3} = 1.383789595 \times 10^{-6}.
\end{equation}
The concatenation of these two rays will be called Ray 3.

%1,8781317501215256 \times 10^{-8} \times 73,679048074068589 =
%138,37895950663845920323077971938 \times 10^{-8}

Further experiments with decreasing $r_b$ were not carried out because at $r_b =
0.005$ a numerical instability, known from previous papers
\cite{Kras2016a,Kras2016b}, showed up: at $\Delta t_c$ slightly larger than
(\ref{7.12}), the past-directed ray overshot the BB hump and hit the BB in the
Friedmann region far from the QSS region, while at $\Delta t_c$ slightly smaller
than (\ref{7.12}), the past-directed ray hit the BB close to $r = 0$ with $1 +
z$ larger than the upper limit in (\ref{7.1}).

%1 + z on the ray up = 73.679048074068589
%1 + z on the ray down = 1,8781317501215256 * 10^{-8}
%73,679048074068589 * 1,8781317501215256 * 10^{-8}
%= 138,37895950663845920323077971938 * 10^{-8}

%Real coordinates of Ray 3 at LSH:
%2,9452138001815902E-003 = r3,
%         -0,13933481010992060 = t3
%Corresponding point on BB has
%t_{Be} = -0,13932954689046649, B = 8,064 * 10^9
%t_B(r3) = t_{Be} - B * (r3)^6 =
%= t_{Be} - 5,2632194658576249566869416062971e-6
%= -0,13932428367100063237504331305839
%t3 - t_{Be} = -0,00000526321945411
%r3^6 = 6,5268098534940785673201160792375e-16
%B*r3^6 = 5,2632194658576249566869416062971e-6
%t3 - t_{Be} + B*r3^6 = 1,174762495668694160629710014347e-14

%\Delta t_{c3} = 0.00000016666400\ {\rm NTU}

%These are the results from szekextredownray5.f90 and szekextreupray5.f90:
%1 + z on the ray down = 7,5237815977402533E-011
%1 + z on the ray up = 114,90149487702448
%7,5237815977402533E-011 * 114,90149487702448 = 8,6449375270860277150014502550078e-9

 \begin{figure}[h]
 \begin{center}
\hspace{-5mm} \includegraphics[scale = 0.64]{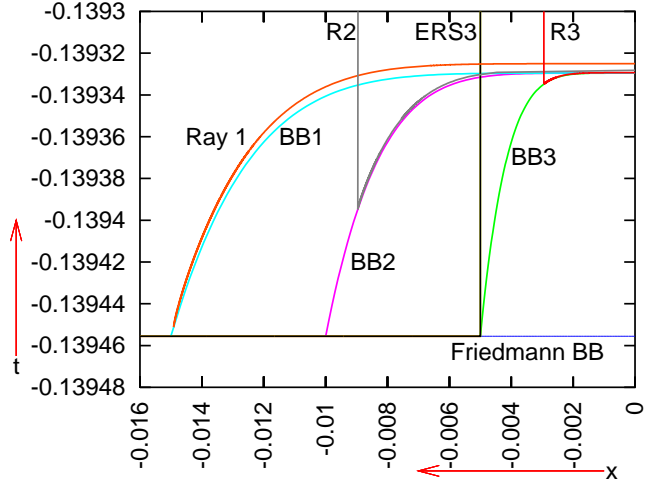}
 \caption{
 \label{down3}
 \footnotesize
The segments of Rays 1 -- 3 between the LSH and $r = 0$, and their corresponding
BB profiles. The coordinate $x = -r$ goes along the dipole maximum. See more
explanation in the text.}
 \end{center}
 \end{figure}

Figure \ref{down3} shows Rays 1, 2 and 3 between $r = 0$ and the LSH, and their
corresponding BB profiles. The dipole maximum is to the left, at $\vartheta =
\pi$. The curves BB1, BB2 and BB3 are the graphs of $t_B(r)$ corresponding to
$r_b = 0.015$, $r_b = r_{b2} = 0.01$ and $r_b = r_{b3} = 0.005$, respectively.
The vertical lines R2 and R3 mark the $x = -r$ coordinates of the points where
Rays 2 and 3, respectively, crossed the LSH. The LSH for each profile is, at the
scale of the figure, indistinguishable from the BB.\footnote{The coordinates of
the point where Ray 3 crossed the LSH are $(r, t) \approx (0.0029452,
-0.1393348\ {\rm NTU})$, while the point on BB3 of the same $r$-coordinate has
its $t$ smaller by $1.175 \times 10^{-14}$ NTU. This is $\approx 10^{-9}$ of the
tics separation in Fig. \ref{down3}. At $r = 0$ the $t$-coordinates of the two
sets differ by $\Delta t_{c3} = 0.00000016666400$ NTU, which is 0.008 of the
tics separation.} The line ERS3 is at the outer edge of the Extremum Redshift
Surface corresponding to BB3. Between $r = 0$ and $r = r_{b3}$, this surface
lies high above the BB hump and nearly horizontally: its $t$-coordinate varies
from 890.8421697 NTU at $r = 0$ to 890.8421435 NTU at $r = r_{b3}$. This is high
above the upper edge of Fig. \ref{down3}. Consequently, all axial rays keep
acquiring blueshift as long as they stay in the QSS region - unlike in Refs.
\cite{Kras2016a,Kras2016b}, where the ERS was tangent to the BB at the origin.
For this reason, a minimum of $\Phi$ generates a stronger blueshift than an
inhomogeneity around the origin, as is seen by comparing (\ref{7.5}),
(\ref{7.11}) and (\ref{7.17}) with $1 + z = 1.553 \times 10^{-5}$ obtained in
Ref. \cite{Kras2018a}.

\section{Decreasing the height of the BB hump}\label{decheight}

\begin{figure}[h]
 \begin{center}
 \includegraphics[scale = 0.6]{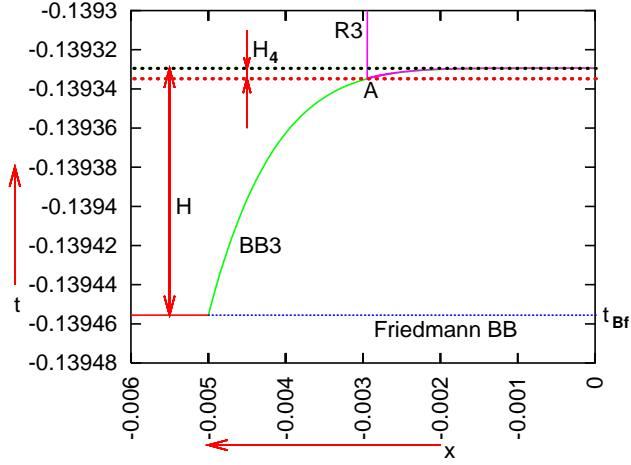}
 \caption{
 \label{only3}
 \footnotesize
The segment of Ray 3 between the LSH and $r = 0$, and the BB3 profile. The
vertical line R3 marks the value of $x = -r$ at which Ray 3 crossed the LSH.
More explanation in the text.}
 \end{center}
 \end{figure}

Figure \ref{only3} shows a closeup view on Ray 3 and BB3 of Fig. \ref{down3}. As
is seen, Ray 3 flew above the BB hump only for about half of the hump's radius;
the remaining part of the inhomogeneity did not influence it. Thus, a stronger
blueshift could be achieved by moving the ray up so that it hits the BB hump
still further down. But our ultimate aim is to give the hump the smallest
possible angular diameter as seen by the present observer. Therefore, in the
next step we lowered the BB hump without changing the ray parameters.

\begin{figure}[h]
 \begin{center}
 \includegraphics[scale = 0.6]{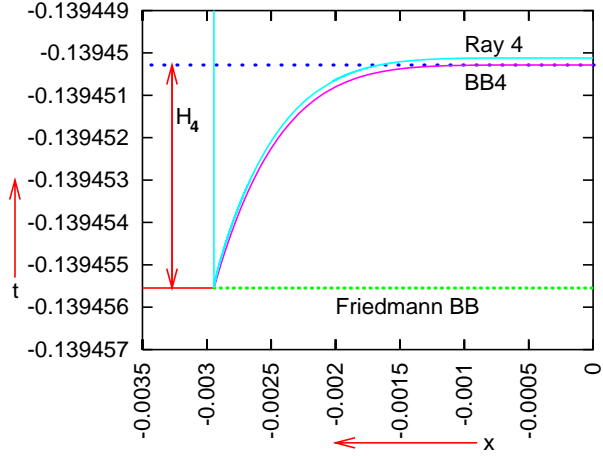}
 \caption{
 \label{down4}
 \footnotesize
The BB profile with the boundary between the QSS and Friedmann regions being at
point A of Fig. \ref{only3}. The vertical straight line marks the value of $x =
-r$ at which the ray crossed the LSH.}
 \end{center}
 \end{figure}

The part of the QSS region to the left of the R3 line did not contribute to the
blueshift on Ray 3, so we replaced it by the Friedmann background. Ray 3 crossed
the LSH at point A in Fig. \ref{only3}, with $(t, r) = (t_A, r_A)$, where
\begin{eqnarray}\label{8.1}
t_A &=& -0.13933481010992060\ {\rm NTU}, \nonumber \\
r_A &=& 2.9452138001815902 \times 10^{-3}.
\end{eqnarray}
The $r_A$ was taken as the new outer boundary of the QSS region, while $t_{\rm
Bf}$ and $B$ of (\ref{5.11}) and (\ref{5.5}) were left the same. After this, the
height of the BB hump decreased from the $H$ of (\ref{5.3}) to
\begin{equation}\label{8.2}
H_4 = t_{\rm Bf} + H - t_A = 5.26321945411 \times 10^{-6}\ {\rm NTU},
\end{equation}
see Fig. \ref{only3}. Then, a past-directed Ray 4 was calculated from the
initial point $(r_4, t_4) \df (0, t_{\rm Bf} + H_4 + \Delta t_{c3})$ with
$\Delta t_{c3}$ as in (\ref{7.12}) along $\vartheta = \pi$. The blueshift on it
on crossing the LSH was
\begin{equation}\label{8.3}
1 + z_{p4} = 1.8786236899437370 \times 10^{-8},
\end{equation}
very close to that of (\ref{7.13}). Figure \ref{down4} shows the corrected BB
configuration and the past-directed part of Ray 4.

%t_{\rm Bf} = -0,13945554689046649, H = 0,000126
%t_A = -0,13933481010992060
%t_{\rm Bf} + H - t_A = 0,00000526321945411

%The future Ray 4 in the paper is Ray 6 from the programs

On the ray propagating from $(r_4, t_4)$ upward to the present time along
$\vartheta = 0$ the $r_b$ parameter had to be changed from $r_{b3}$ to $r_A$.
The redshift on it between $r = 0$ and the present time came out to be
\begin{equation}\label{8.4}
1 + z_{f4} = 458.91884554506117.
\end{equation}
Consequently, the total $1 + z$ between the LSH and the present time was
\begin{equation}\label{8.5}
1 + z_{t4} = \left(1 + z_{p4}\right) \left(1 + z_{f4}\right) = 0.862135815
\times 10^{-5}.
\end{equation}
%full value calculated by WinEdt calculator:
%1,8786236899437370 \times 10^{-8} \times 458,91884554506117 =
%862,13581500258272545421552339229 \times 10^{-8}
%= 0,86213581500258272545421552339229 \times 10^{-5}
This is safely within the range defined by (\ref{7.1}). This $z_{t4}$ was
achieved with the radius of the BB hump (as measured by $r$) and its height
$H_4$ being 0.196 and 0.042, respectively, of those in Ref. \cite{Kras2018b}.

In consequence of numerical inaccuracies, the future endpoint of Ray 4 overshot
the present time $t = 0$. The coordinates of the endpoint were
\begin{eqnarray}
t_{\rm now4} &=& 7.6253109886207342 \times 10^{-11}\ {\rm NTU},\ \ \ \
\label{8.6} \\
r_{\rm now4} &=& 0.95434899416269714. \label{8.7}
\end{eqnarray}

\begin{figure}[h]
 \begin{center}
\hspace{-8mm} \includegraphics[scale = 0.65]{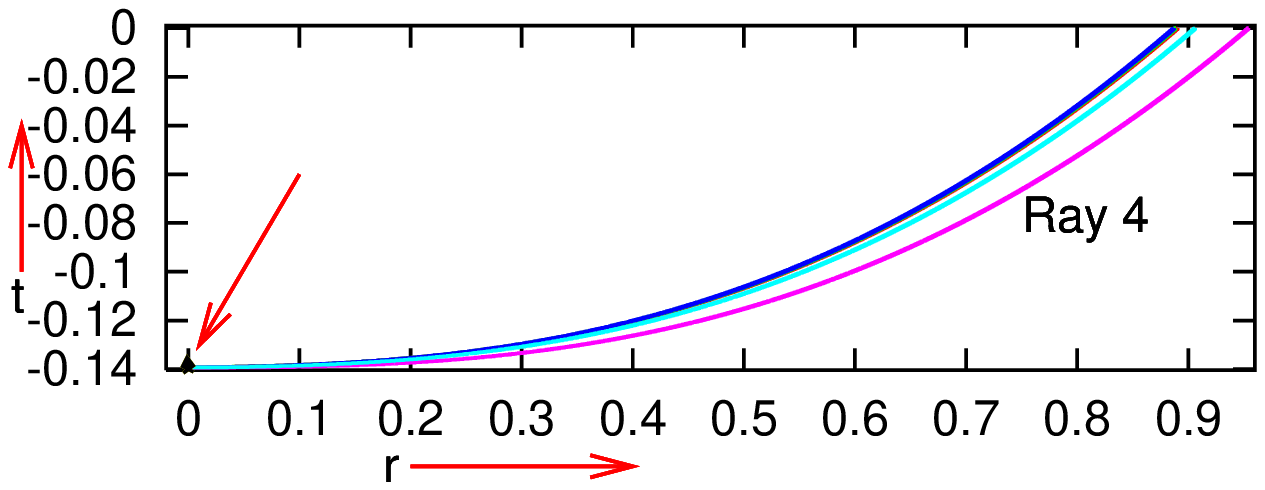}
 \hspace{-1.5cm}
 ${ }$ \\[5mm]
\includegraphics[scale = 0.65]{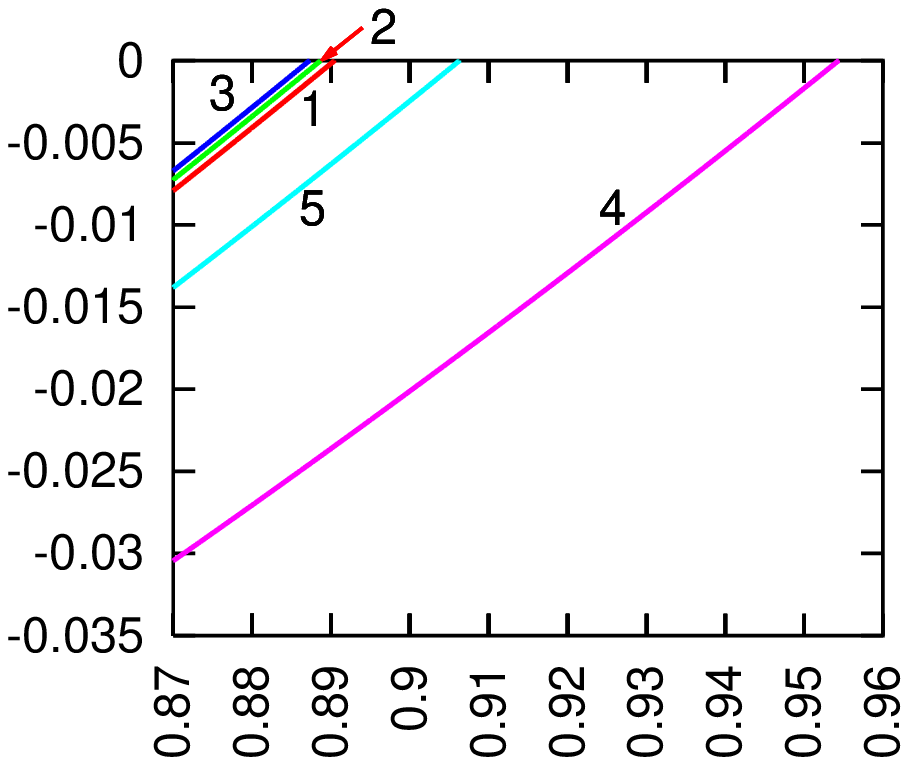}
% ${ }$ \\[-1cm]
 \caption{
 \label{upcaly}
 \footnotesize
{\bf Upper panel:} Rays 1 -- 5 shown from the LSH to the present time. The arrow
points to the graph of the BB3 profile, which is the tiny dot. {\bf Lower
panel:} The same rays near their upper ends. The difference in $r$ between the
center of the BB hump and the observer is largest for Ray 4 and smallest for Ray
3.}
 \end{center}
 \end{figure}

For completeness, a similar operation to that described above was done on the
BB2 profile. The QSS/Friedmann boundary was moved from $r_b = r_{b2} = 0.01$ to
$r = r_{b5}$, slightly beyond the $r$ at which Ray 2 crossed the LSH:
\begin{equation}\label{8.8}
r_{b5} = 0.0090765667.
\end{equation}
The corresponding $t$ on BB2 is
\begin{equation}\label{8.9}
t_{b5} = -0.1394\ {\rm NTU}.
\end{equation}
This resulted in replacing the $H$ of (\ref{5.3}) by
\begin{equation}\label{8.10}
H_5 = t_{\rm Bf} + H - t_{b5} = 7.04531 \times 10^{-5}\ {\rm NTU}.
\end{equation}
The ray sent to the future from $(r, t) = (0, t_{\rm Bf} + H_5 + \Delta t_{c2})$
(the same $\Delta t_{c2}$ as in (\ref{7.6})) is Ray 5 from Fig. \ref{upcaly}. As
the other rays, it overshot the present time by $t_{\rm now5}$, and the
parameters of the endpoint were
\begin{eqnarray}
1 + z_{f5} &=& 84.123779615683631, \label{8.11} \\
t_{\rm now5} &=& 8.6312831305632174 \times 10^{-10}\ {\rm NTU},\ \ \ \  \label{8.12} \\
r_{\rm now5} &=& 0.90628860720677851. \label{8.13}
\end{eqnarray}
The total $1 + z$ between the LSH and $t_{\rm now5}$ was thus, from (\ref{7.7})
and (\ref{8.11}),
\begin{equation}\label{8.14}
1 + z_{t5} = \left(1 + z_{p2}\right) \left(1 + z_{f5}\right) = 6.32928945 \times
10^{-9}.
\end{equation}
This is much better than the lower end of the range (\ref{7.1}).

%t_{\rm Bf} = -0,13945554689046649, H = 0,000126
%t_{b5} = -0,1394
%H_5 = t_{\rm Bf} + H - t_{b5} =

% real r_{b5} = 9.0765667005013265E-003
% real H_5 = 7.04531095334915e-05
% both calculated by gnuplot
% 1 + z_{f5} and  (r, t)_{now} calculated in the Fortran program
% real 1 + z_{t5} = 7,5237815977402533 \times 10^{-11} * 84,123779615683631
%= 6,3292894500483714096092124660373 \times 10^{-9}
% calculated by the WinEdt calculator

 \begin{figure}[h]
 \begin{center}
\hspace{-8mm} \includegraphics[scale = 0.55]{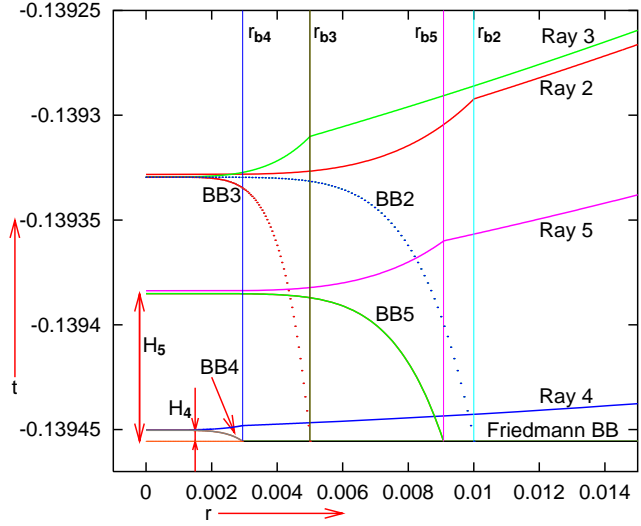}
 \caption{
 \label{athump}
 \footnotesize
The segments of Rays 2 -- 5 going from $r = 0$ toward the present time, shown in
and near the QSS region, and their corresponding BB profiles.}
 \end{center}
 ${ }$ \\[-1cm]
 \end{figure}

Figure \ref{upcaly} shows the $t(r)$ graphs of Rays 1 -- 5 all along their
length (the upper panel) and near their upper ends (the lower panel).

Figure \ref{athump} shows the segments of Rays 2 -- 4 between $r = 0$ and $r =
0.015$, and their corresponding BB profiles. Between the LSH and $r = 0$, Ray 3
has the same shape as Ray 4 and would coincide with it when translated down by
$H - H_4$. Similarly, Ray 2 would coincide with Ray 5 between the LSH and $r =
0$ when translated down by $H - H_5$. The same is true for the pairs of BB
profiles (BB4, BB3) and (BB5, BB2).

\section{Tracing the rays back from the present time}\label{backfromnow}

In the next section we will calculate the angular radius of the QSS region
corresponding to BB4 as seen by the observer at $t = 0$ who receives the
maximally blueshifted gamma ray. For this purpose, we will have to integrate
(\ref{3.8}) -- (\ref{3.12}) backward in time from the observer position and find
the ray that grazes the boundary of the QSS region. But we must verify whether
the observer position was correctly identified, i.e., whether the axial ray
emitted from the endpoint of Ray 4 at $t = 0$ toward the past coincides with Ray
4 at $r = 0$. As will be seen below, it does not: the two rays nearly coincide
between $t = 0$ and the QSS/Friedmann boundary, but the backward ray (hereafter
called IR 4, short for ``inverse Ray 4'') enters the QSS region with a different
$\dril t r$ than Ray 4 had on leaving it. This problem, caused by numerical
inaccuracies, existed also in Refs. \cite{Kras2018a,Kras2018b}. The present
section explains how this discrepancy was handled.

 \begin{figure}[h]
 \begin{center}
 \includegraphics[scale = 0.55]{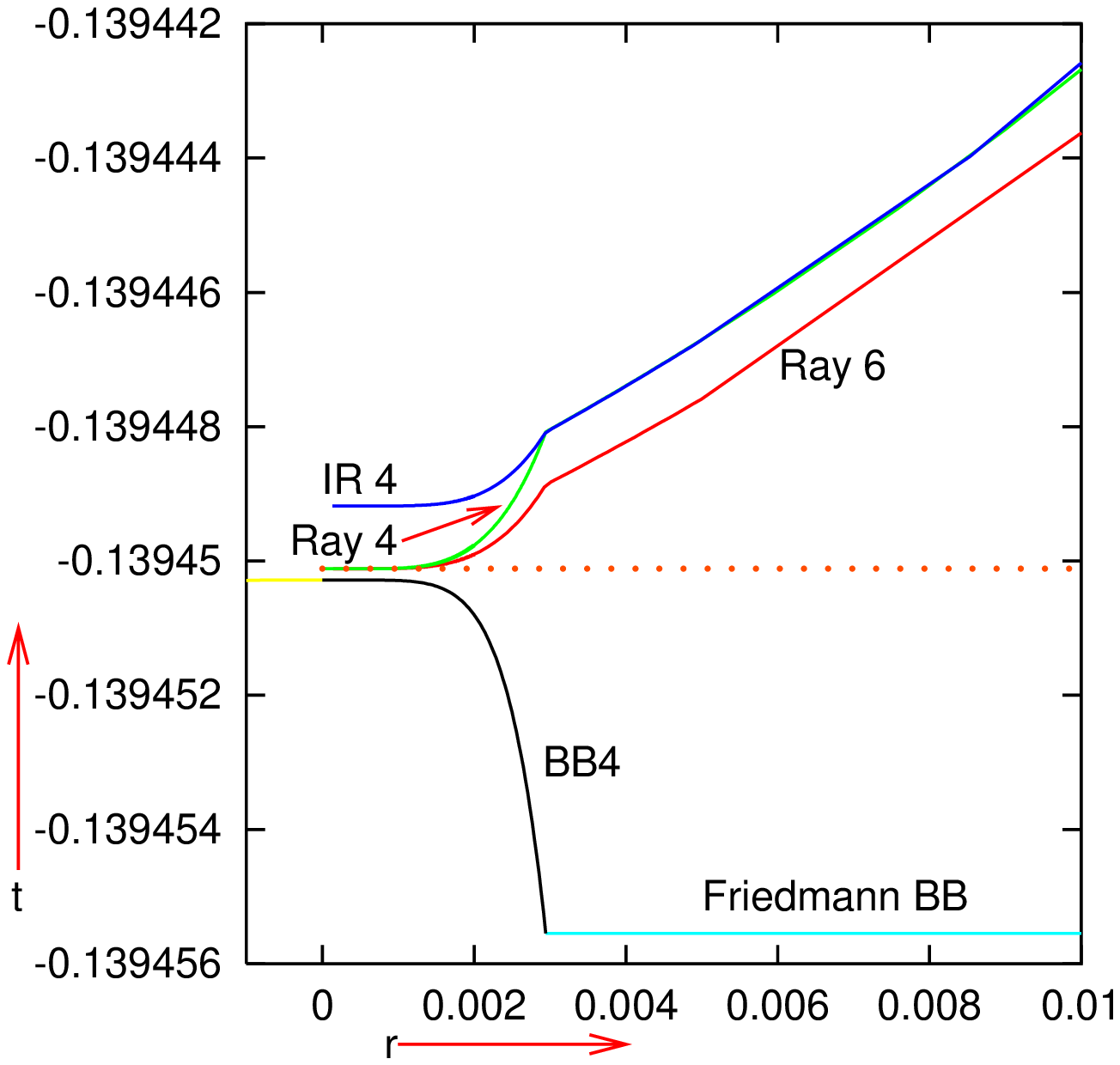}
 \includegraphics[scale = 0.55]{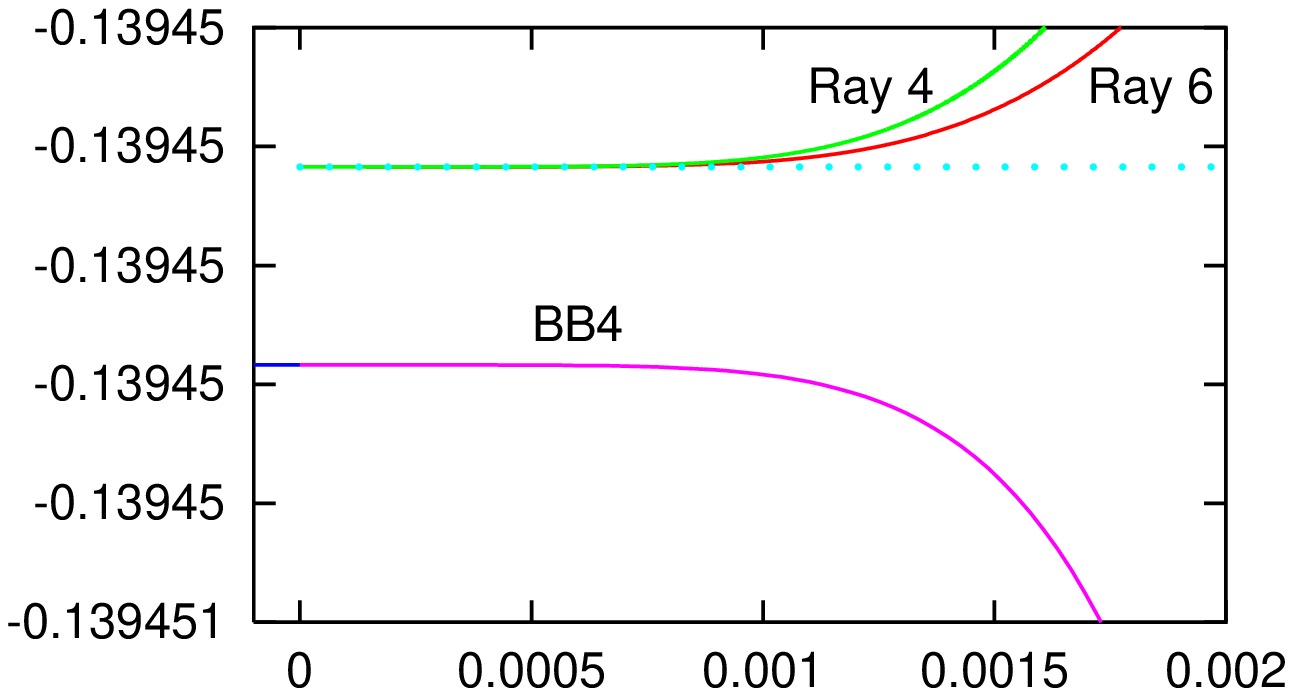}
 \caption{
 \label{drawzbokudol}
 \footnotesize
{\bf Upper panel:} Rays 4, 6 and IR 4 in a vicinity of the BB4 hump. The
dotted line marks the $t$-coordinate of Ray 4 at $r = 0$. The
difference between this $t$ and the top of BB4 is the $\Delta t_{c3}$ of
(\ref{7.12}). The $t$-coordinates of Rays 6 and 4 at $r = 0$ differ by 0.12\% of
$\Delta t_{c3}$, see text. {\bf Lower panel:} A closeup view on the
neighbourhood of $(t, r) = (t_B(0) + \Delta t_{c3}, 0)$ in the upper panel. The
difference between Rays 4 and 6 at $r = 0$ is not visible at the scale of this
figure. Ray IR 4 is above the upper edge of the figure.}
 \end{center}
 \end{figure}

 \begin{figure}[h]
 \begin{center}
 \includegraphics[scale = 0.55]{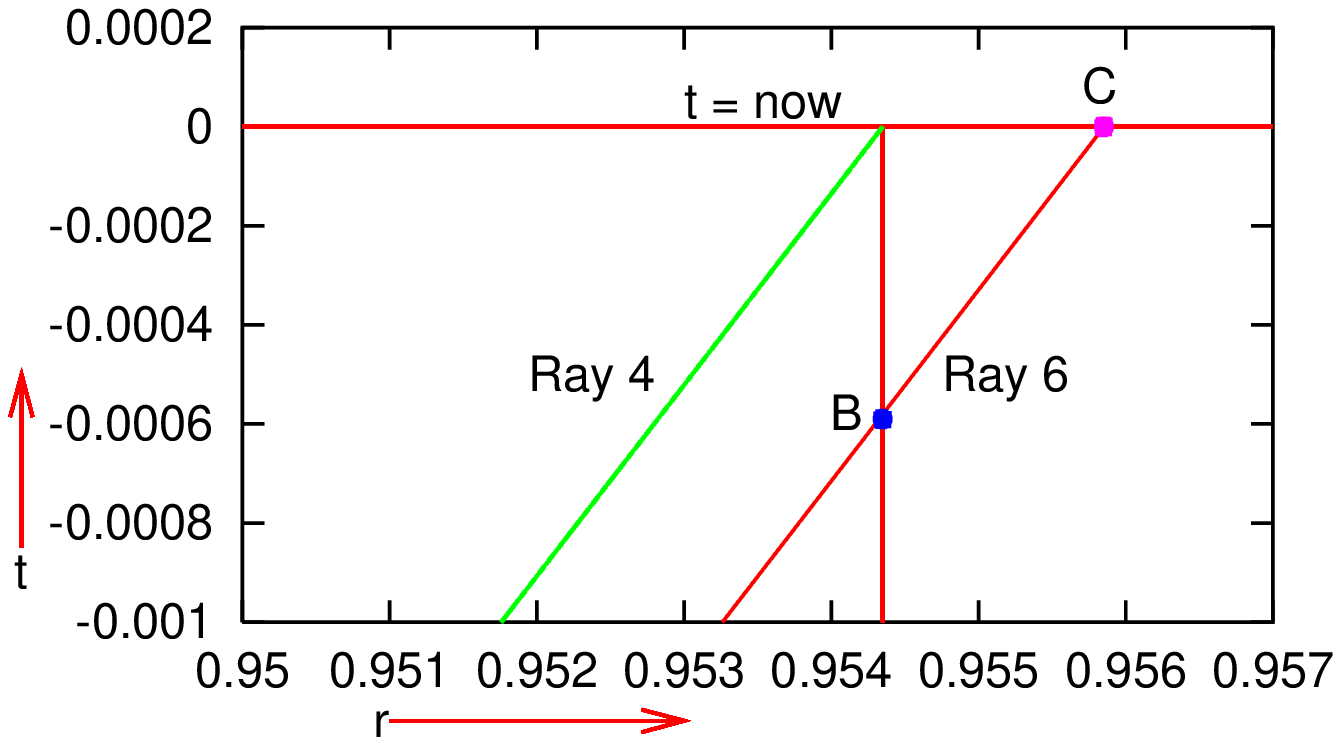}
 \caption{
 \label{drawzbokugor}
 \footnotesize
Rays 4 and 6 in a vicinity of the present time $t = 0$. The difference between
$t = 0$ and the actual values of $t$ on the upper ends of the two rays is not
visible at the scale of the figure. The vertical line marks the $r$-coordinate
on Ray 4 at the present time. The meaning of points B and C is explained in
Appendix \ref{ray6}.}
 \end{center}
 \end{figure}

The IR 4 was sent from $(t, r) = (t_{\rm now4}, r_{\rm now4})$ given by
(\ref{8.6}) -- (\ref{8.7}), and arrived at $r = 0$ with $t$ differing visibly
from that of Ray 4, see the upper panel of Fig. \ref{drawzbokudol}. The $t(0) -
t_B(0)$ on IR 4 was $\approx 6.6 \times \Delta t_{c3}$ instead of $\Delta
t_{c3}$ for Ray 4 given by (\ref{7.12}). So, the initial point of the
past-directed ray was hand-corrected so as to achieve a better coincidence at $r
= 0$. On Ray 6 shown in Fig. \ref{drawzbokudol}, the ratio $(t(0) - t_B(0)) /
\Delta t_{c3}$ was $\approx 0.9988$, and it was taken to be a satisfactory
precision. The initial point of Ray 6 is at
\begin{eqnarray} \label{9.1}
t_{\rm now6} &=& 1.9143125092526522 \times 10^{-11}\ {\rm NTU},\ \ \ \ \nonumber
 \\
r_{\rm now6} &=& 0.95585224106471711.
\end{eqnarray}
Appendix \ref{ray6} explains how this point was found. The $1 + z$ on Ray 6
between the point of coordinates (\ref{9.1}) and $(t(0), 0)$ was
568.65551516257369 -- rather strongly off the value (\ref{8.4}), but this
discrepancy has no influence on the calculation of the angular radius in the
next section. The real redshift along this geodesic segment should be between
these values. Figure \ref{drawzbokugor} shows Rays 4 and 6 in a vicinity of the
present time $t = 0$. The real $r$-coordinate of the observer receiving the ray
with the strongest blueshift should be between $r_{\rm now4}$ of (\ref{8.7})
$r_{\rm now6}$ of (\ref{9.1}). We will calculate the angular radius of the light
source for both these positions of the observer.

%Real value of $(t(0) - t_B(0))/\Delta t_{c3}$ on Ray 6: 0.99883535807315638
% 1 - (the above) = 0,00116464192684362

See Appendix \ref{numprec} for remarks on numerical precision.

\section{The angular size of the source of the blueshifted rays}\label{angrad}

\begin{figure}[h]
 \begin{center}
 \includegraphics[scale = 0.55]{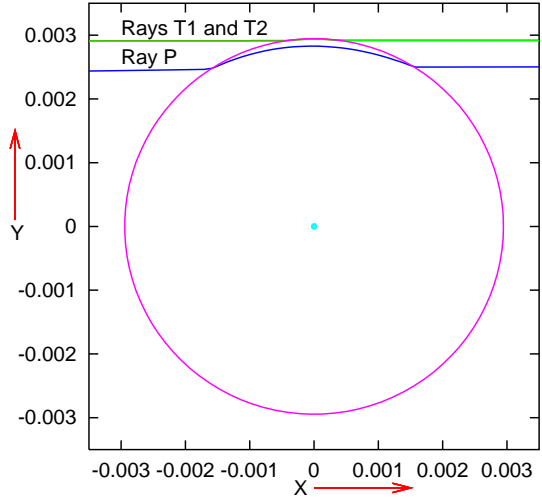}
 \caption{
 \label{drawgrazerayforpub}
\footnotesize Projection of the rays that graze the QSS region on a surface of
constant $t$ along the flow lines of the dust filling the spacetime. The
boundary of the QSS region is the large circle. Rays T1 and T2 correspond to the
two positions of the observer described in the text; at the scale of this figure
they coincide. Ray P is a projection of an exemplary ray that penetrates the QSS
region. The coordinates in the figure are $(X, Y) = r (\cos \vartheta, \sin
\vartheta)$, with $(r, \vartheta)$ being those of (\ref{3.8}) -- (\ref{3.12}). }
 \end{center}
 \end{figure}

\begin{figure}[h]
 \begin{center}
 \includegraphics[scale = 0.6]{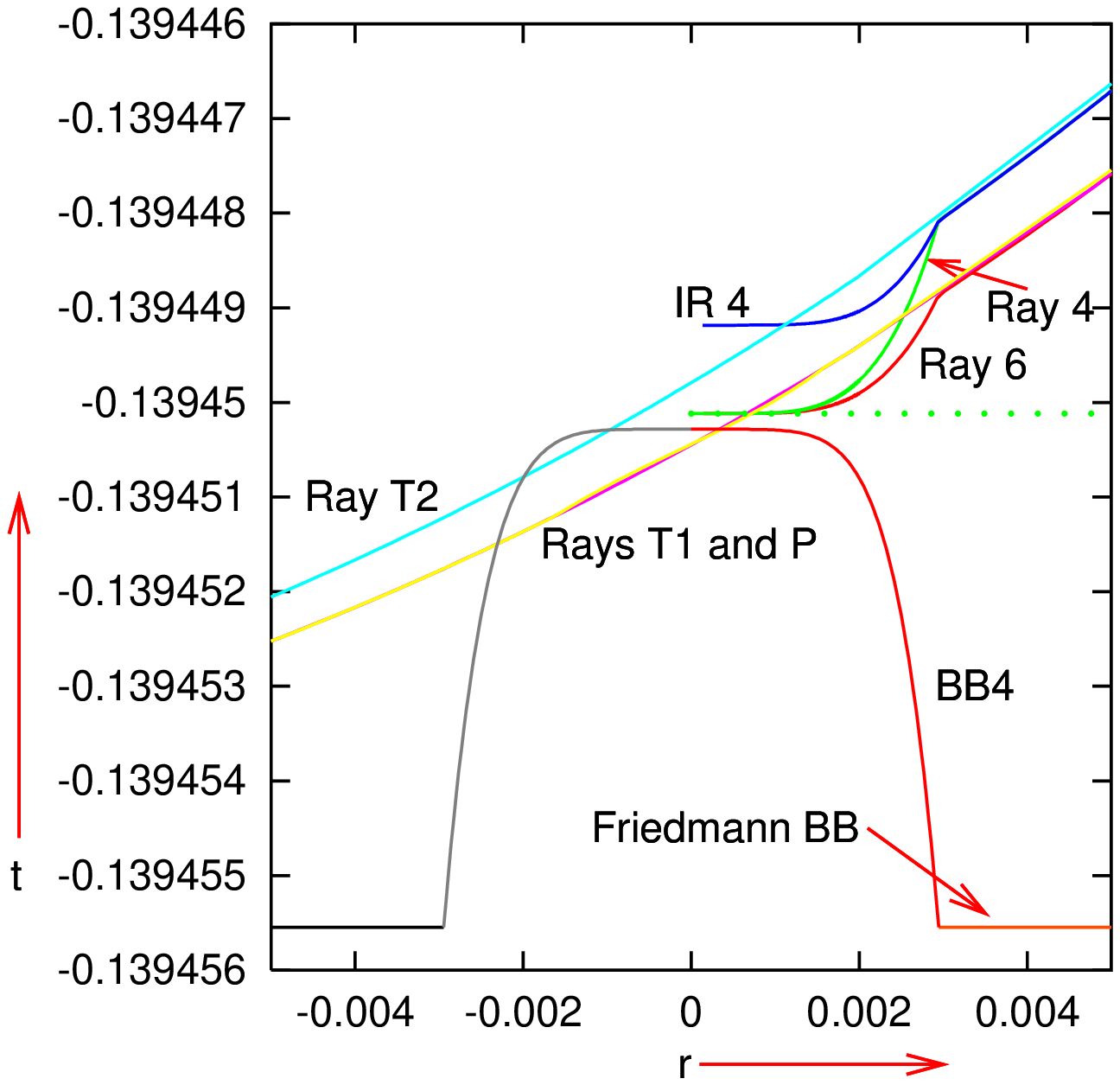}
 \caption{
 \label{drawzbokudolforpub}
\footnotesize This is a modified version of Fig. \ref{drawzbokudol}, which
includes the rays shown in Fig. \ref{drawgrazerayforpub}. The nonaxial rays T1,
T2 and P are shown here projected on the $Y = 0$ surface along lines of constant
$t$ and $X$.}
 \end{center}
 \end{figure}

To determine the angular radius of the QSS region seen by a present observer one
has to shoot a past-directed ray from the observer position in such a direction
that it grazes the boundary of the inhomogeneity, call it Ray T. This ray was
found by trial and error. Then the angle $\alpha$ between Ray T and the axial
ray (the one that passes through $r = 0$) is the desired angular radius. As
shown in Ref. \cite{Kras2018a} it is given by
\begin{equation}\label{10.1}
\cos \alpha = \sqrt{1 - \left(k^{\vartheta}_{o} \Phi_o\right)^2} \Longrightarrow
 \sin \alpha = k^{\vartheta}_{o} \Phi_o,
\end{equation}
where $k^{\vartheta}_{o}$ is the $\vartheta$ component of the vector
$k^{\alpha}$ tangent to Ray T at the observer and $\Phi_o$ is the value of the
metric function $\Phi$ at the observer. This calculation was done for two
observer positions: the initial point of Ray 6 given by (\ref{9.1}) and the
endpoint of Ray 4 given by (\ref{8.6}) -- (\ref{8.7}). The difference is not
significant: the angular radius for the first observer is
\begin{equation}\label{10.2}
\alpha_1 = 0.00308221\ {\rm rad} = 0.1765976^{\circ},
\end{equation}
and for the second observer it is
\begin{equation}\label{10.3}
\alpha_2 = 0.0030774\ {\rm rad} = 0.1763199^{\circ};
\end{equation}
the corresponding rays are denoted T1 and T2 in Figs. \ref{drawgrazerayforpub}
and \ref{drawzbokudolforpub}. In Ref. \cite{Kras2018a}, the angular radius of
the QSS region around the origin was between 0.96767$^{\circ}$ and
0.9681$^{\circ}$, depending on the direction of observation. Whichever
combination of two radii we take, the ratio of the radius found here to that in
Ref. \cite{Kras2018a} is $\approx 0.182$. The difference between (\ref{10.2})
and (\ref{10.3}) is influenced by the numerical error in determining the impact
parameter of the ray relative to $r = 0$. For the first observer this parameter
is $0.9976 \times r_b$, for the second one it is $0.9968 \times r_b$. These
numbers show that the ``grazing'' rays actually entered the QSS region a little.
However, the redshift on them between the LSH and the present time does not
significantly differ from that on the ray that stayed in the Friedmann region
all the way. On two all-Friedmannian rays reaching the first observer, $1 + z$
was
\begin{equation}\label{10.4}
951.55845651643119 \quad {\rm and} \quad 951.56113626862839,
\end{equation}
while on the ``grazing'' rays the respective values were 951.56298581163151 and
951.63204672978486. On Ray P, for which the impact parameter was $0.96 \times
r_b$ the redshift was $1 + z = 1026.4529080967900$, i.e., $z$ was larger than on
the grazing rays. This is consistent with what was found in Ref.
\cite{Kras2018a}: on decreasing the impact parameter from the edge of the QSS
region, $z$ at first increased above the background value before it started to
decrease. Figures \ref{drawgrazerayforpub} and \ref{drawzbokudolforpub} show
Rays T1, T2 and P in two views.\footnote{The all-Friedmannian rays referred to
in (\ref{10.4}) are beyond the margins of Fig. \ref{drawgrazerayforpub}. They
crossed the LSH at $(X, Y) = (-0.00046394, 0.00375996)$ and $(-0.00046585,
0.003339)$, respectively.}

The angular radii (\ref{10.2}) and (\ref{10.3}) are smaller than the angular
resolution for most of the 186 GRBs detected by the LAT between 2008 and 2018
\cite{Ajel2019}: the localisation error was smaller than 0.18$^{\circ}$ in 55
cases.

An interesting question now is: how many circles of angular radius $\alpha$ can
be placed on the celestial sphere without overlapping? A method to tackle this
question was suggested in Ref. \cite{Kras2018a}. We imagine each circle being
inscribed into a quadrangle of arcs of great circles on a sphere $S_c$ of radius
$R_c$, and then divide the surface area of $S_c$ by the surface area of the
quadrangle. The resulting number $N$ is only an approximate estimate because
such shapes cannot completely cover the sphere: the quadrangles will leave holes
between them. However, this method takes into account some of the area outside
the circles, so it yields a better approximation than dividing $4 \pi {R_c}^2$
by the surface area of the small circle.\footnote{The actual number is lower
than the one in (\ref{10.6}) because this method assumes that the holes between
quadrangles were also covered.} By Ref. \cite{Kras2018a},
\begin{equation}\label{10.5}
N = \frac {\pi} {\arcsin \left(\sin^2 \alpha\right)}.
\end{equation}
Taking $\alpha = 0.00308221$ rad, the result is
\begin{equation}\label{10.6}
N = 330\ 694,
\end{equation}
which is $\approx 30$ times the number for QSS regions that contain an origin
\cite{Kras2018a}.

%exactly N = 330 694,36730208196379957724618433

%real angular radii: 0.0030822103526155706\ {\rm rad} = 0.17659764477640147^{\circ}
%and 0.0030773630166093572\ {\rm rad} = 0.17631991288136364^{\circ}
%Real values $0.99763344881791616 \times r_b$ and $0.99681726986060304 \times r_b$

%smallest angular radius in Ref. \cite{Kras2018a}: 0,96767,
% 0,96767/0,17628792893241552 = 5,4891449792400760977034981120217
% 0,96767/0,17631991288136364 = 5,4881492633852085200649028034711

\section{Conclusion}\label{sumcon}

In the previous papers \cite{Kras2018a} -- \cite{Kras2018b}, QSS regions
possessing origins were employed to consider the same process as the one
considered here: matter inhomogeneities blueshifting (along preferred
directions) rays of the relic radiation from their initial frequencies to the
gamma range. The conclusion of the present paper is: when the QSS region does
not possess an origin, but surrounds a spatial minimum of the areal radius
function $\Phi$, then it may be a few times smaller in diameter and its
amplitude of $t_B(r)$ may be several times lower, and yet it will generate gamma
rays of the same frequency range. The angular radius of the gamma-ray source
seen by the present observer is here between $0.176^{\circ}$ and
$0.177^{\circ}$, which is $\approx 0.182 < 1/5$ of that in the previous papers.
The amplitude of the bang-time function $t_B(r)$ (the $H_4$ in (\ref{8.2})) is
here $\approx 0.042 < 1/23$ of that in Refs. \cite{Kras2018a} --
\cite{Kras2018b}. The reason of the improvement is that the extremum redshift
surface is tangent to the BB at an origin (the case considered in the former
papers), but is not tangent to it at the minimum of $\Phi$ (the case considered
here). Consequently, in the present case light rays passing through the QSS
region spend more time in the blueshift-generating zone. This is why a smaller
inhomogeneity around a minimum of $\Phi$ is needed to generate the same range of
blueshift.

It must be strongly emphasised that $\alpha_1$ and $\alpha_2$ given in
(\ref{10.2}) and (\ref{10.3}) are NOT the lower bounds on the angular radii of
sources of gamma rays. The inhomogeneity that produced these numbers is an
example -- a proof of existence of a sufficiently small source of the gamma
radiation, and no optimisation was attempted. So, it must be possible to make it
still smaller. It would be incredible to find the absolute minimum of diameter
and amplitude by blind search -- and the same is true for the configurations
considered in Ref. \cite{Kras2018a}. Thus, there is room for further
improvements (for example by allowing the $E(r)$ function to be
non-Friedmannian).

Depending on the shape of the $t_B(r)$ function, the rays emitted from the
last-scattering hypersurface as hydrogen and helium emission radiation may be
blueshifted to different bands, not necessarily to the gamma-ray frequencies.
For example, they may end up reaching the present observers as X- or ultraviolet
rays. In the latter cases the required blueshifts would be weaker ($z$ would not
have to be as close to $-1$ as in (\ref{7.1})), so lower and narrower humps on
the BB would suffice. Consequently, reconciling these inhomogeneities with the
observed limits on anisotropies of the CMB radiation (directional temperature
differences $\Delta T/T \approx 10^{-5}$) would be easier. The reason why Refs.
\cite{Kras2016a} -- \cite{Kras2018b} and the present paper concentrated on
blueshifting to the gamma range is just because this is the most difficult case.
This author did not wish to be suspected of choosing the easy ways.

The papers \cite{Kras2016a} -- \cite{Kras2018b} and the present one discussed
only the conditions for blueshifting the initial frequencies to the GRB range.
The questions of the expected present intensity of the blueshifted radiation and
of its spectrum were not considered, and the answers to them are crucial for the
problem of detectability. This will have to be dealt with in the future; it
might happen that in our real Universe the signal is too weak to be detected at
present. But the right time to consider detectability will come when we clearly
understand what kind of signal should be expected, and this is what the papers
were meant to clarify.

\appendix

\section{When do ERS and BB coincide at the origin?}\label{ERSatorig}

Equations (\ref{6.4}) and (\ref{6.5}) were derived without using any explicit
choice of the $r$-coordinate (no use was made of (\ref{4.7}) -- (\ref{4.9})).
So, in this appendix we can choose $r$ so that $r = 0$ at the origin, not at an
extremum, and $M = M_0 r^3$. We recall that with such choice of $r$, and with $P
= Q = 0$ and $S$ given by (\ref{4.10}) the origin is nonsingular when
\cite{HeKr2002,PlKr2006}
\begin{equation}\label{a.1}
E = - \tfrac 1 2\ k r^2 + {\cal O}_2(r) \qquad \left(\Longrightarrow E,_r = - k
r + {\cal O}_1(r)\right),
\end{equation}
where ${\cal O}_{\ell}(r)$ denotes a function that has the property
\begin{equation}\label{a.2}
\lim_{r \to 0} \frac {{\cal O}_{\ell}(r)} {r^{\ell}} = 0
\end{equation}
for $\ell \geq 0$ (with $\ell = 0$ this means $\lim_{r \to 0} {\cal O}_0(r) =
0$). No approximations will be used along the way -- the whole calculation will
be exact, but the explicit forms of the functions hidden in ${\cal O}_{\ell}(r)$
will be irrelevant.

Substituting $M = M_0 r^3$ and (\ref{a.1}) in (\ref{6.5}) we obtain
\begin{eqnarray}\label{a.3}
&& \left[\left(\frac {- k r + {\cal O}_1(r)} {- k r^2 + 2 {\cal O}_2(r)} - \frac
1 2\ \frac {\varepsilon n r^{n - 1}} {r^n + a^2}\right) \sinh \eta \cosh
\eta\right. \nonumber \\
&& + \left(\frac 3 r - 2 \frac {- k r + {\cal O}_1(r)} {- \tfrac 1 2 k r^2 +
{\cal O}_2(r)} + \frac 1 2\ \frac {\varepsilon n r^{n - 1}} {r^n + a^2}\right)
\sinh \eta \nonumber \\
&& + \left.\left(\frac 3 2\ \frac {- k r + {\cal O}_1(r)} {- \tfrac 1 2 k r^2 +
{\cal O}_2(r)} - \frac 3 r\right) \eta\right] \times \sqrt{- k r^2 + 2 {\cal
O}_2(r)} \nonumber \\
&& + \frac {\left(- k r^2 + 2 {\cal O}_2(r)\right)^2} {M_0 r^ 3}\ \dr {t_B} r =
0.
\end{eqnarray}
Now we factor out $r$ from $\sqrt{- k r^2 + 2 {\cal O}_2(r)}$ and multiply by
$r$ each term in the long square bracket. We note that $\lim_{r \to 0} {\cal
O}_2 / r^2 = \lim_{r \to 0} {\cal O}_1 / r = 0$, so
\begin{eqnarray}
&& \lim_{r \to 0} \left(\frac {- k r^2 + r {\cal O}_1(r)} {- k r^2 + 2 {\cal
O}_2(r)}\right) = 1, \label{a.4} \\
&& \lim_{r \to 0} \left[\frac {\left(- k r^2 + 2 {\cal O}_2(r)\right)^{3/2}}
{M_0 r^ 3}\right] = \frac {(-k )^{3/2}} {M_0}. \label{a.5}
\end{eqnarray}
Then, in the limit $r \to 0$, (\ref{a.3}) becomes
\begin{equation}\label{a.6}
4 \sinh^3 (\eta_0 / 2) \cosh (\eta_0 / 2) + \frac {(-k )^{3/2}} {M_0}\ \lim_{r
\to 0} \left(r\ \dr {t_B} r\right) = 0,
\end{equation}
where $\eta_0 = \lim_{r \to 0} \eta$. This shows that $\eta_0 = 0$ (i. e., the
ERS and BB coincide at the origin) if and only if $\lim_{r \to 0} \left(r \dril
{t_B} r\right) = 0$. $\square$

\section{Solvability of Eq. (\ref{6.6})}\label{solvable}

The second line of (\ref{6.8}) shows that when $\varepsilon = +1$, $F_1(r) > 0$
in consequence of (\ref{4.20}). When $\varepsilon = -1$, $F_1(r) > 0$ in
consequence of $A > 0$ and $E_{\rm ext} > 0$, see the comment under
(\ref{4.10}).

{}From (\ref{6.7}) we see that $\left.{\cal H}\right|_{\eta = 0} = 0$. Further
\begin{equation}\label{b.1}
\pdr {\cal H} {\eta} = 2F_1 \cosh^2 \eta + F_2 \cosh \eta + F_3 - F_1.
\end{equation}
{}From here,
\begin{eqnarray}
\left.\pdr {\cal H} {\eta}\right|_{\eta = 0} &=& F_1 + F_2 + F_3 = 0, \label{b.2} \\
\pdr {^2 {\cal H}} {\eta^2} &=& \sinh \eta\ \left(4F_1 \cosh \eta + F_2\right).
\label{b.3}
\end{eqnarray}
Now we define
\begin{equation}\label{b.4}
{\cal G}(r, \eta) \df 4F_1 \cosh \eta + F_2,
\end{equation}
and find, using (\ref{6.8}) and (\ref{6.9})
\begin{equation}\label{b.5}
\left.{\cal G}\right|_{\eta = 0} = 4F_1 + F_2 = \frac {2Da^n + (2 - 3
\varepsilon) Dr^n - 3 \varepsilon M_{\rm ext}} {\left(M_{\rm ext} + Dr^n\right)
\left(r^n + a^n\right)}.
\end{equation}
When $\varepsilon = -1$, this is obviously positive in consequence of $D$ and
$M_{\rm ext}$ being positive, see the comment under (\ref{4.10}). When
$\varepsilon = +1$, this is positive in consequence of the first of
(\ref{4.14}), so
\begin{eqnarray}
&& {\cal G}(0) > 0, \label{b.6} \\
&& \pdr {\cal G} {\eta} = 4F_1 \sinh \eta, \label{b.7}
\end{eqnarray}
which is positive for all $\eta > 0$ in consequence of $F_1(r) > 0$.

Consequently, ${\cal G}(r, \eta) > 0$ for all $\eta \geq 0$, so $\pdril {^2
{\cal H}} {\eta^2} > 0$ for all $\eta > 0$. Then, from (\ref{b.2}), $\pdril
{\cal H} {\eta} > 0$ for all $\eta > 0$. Since $\left.{\cal H}\right|_{\eta = 0}
= 0$, this means ${\cal H} > 0$ for all $\eta > 0$.

The numerator of $F_3$ is $F_{3{\rm n}} = A \left(3M_{\rm ext} + Dr^n\right) -
2DE_{\rm ext} \leq A \left(3M_{\rm ext} + D{r_b}^n\right) - 2DE_{\rm ext}$ since
$r \leq r_b$. Substituting for $M_{\rm ext}$ from (\ref{4.19}) and for $E_{\rm
ext}$ from (\ref{4.20}), we obtain $F_{3{\rm n}} \leq 3AM_0 {r_b}^3 + Dk
{r_b}^2$. With the values of $A$, $M_0$, $r_b$, $D$ and $k$ given in
(\ref{5.7}), (\ref{4.17}), (\ref{5.2}), (\ref{5.9}) and (\ref{4.16}), $3AM_0
{r_b}^3 + Dk {r_b}^2 < 0$, so $F_3 < 0$ ((\ref{4.14}) alone did not guarantee
this).

To find an initial $\eta$ for a numerical program solving (\ref{6.6}), we use
(\ref{b.2}) to write (\ref{6.7}) in the form
\begin{equation}\label{b.8}
{\cal H} = F_1(r) \sinh \eta (\cosh \eta - 1) - F_3(r) (\sinh \eta - \eta).
\end{equation}
Now we observe that, for all $\eta > 0$,
\begin{eqnarray}\label{b.9}
&& \cosh \eta - 1 > \eta^2 / 2, \qquad \sinh \eta > \eta, \nonumber \\
&& \sinh \eta - \eta > \eta^3 / 6.
\end{eqnarray}
Since $F_3 < 0$, (\ref{b.8}) and (\ref{b.9}) imply that for all $\eta > 0$,
\begin{equation}\label{b.10}
{\cal H} > \left(F_1 / 2 - F_3 / 6\right) \eta^3 \df {\cal H}_i.
\end{equation}
Hence, every $\eta$ that solves (\ref{6.6}) is smaller than the $\eta_i$ that
solves ${\cal H}_i = F_4(r)$. Thus, $\eta_i$ can be used as the initial upper
limit on $\eta$ in solving (\ref{6.6}) by the bisection method. The lower limit
is $\eta = 0$ since we showed that $F_4(r) > 0$ for all $r$, while ${\cal H} =
0$ at $\eta = 0$.

\section{Determining the upper end of Ray 6}\label{ray6}

%\setcounter{equation}{0}

%The numbers in this appendix do not agree with those in the programs of the
%'downfromnow' series.
%In reality, the operation described below was done not on the final Ray 4,
%which is upray 6 in the programs, but on the original upray 4, so the
%t-coordinate of point B in Fig. \ref{drawzbokugor} differs from the one on
%Ray 6 in the figure (which was downfromnow3 in the programs).
%But this appendix is only descriptive, so the exact values do not matter.

Since the IR 4 ray reached $r = 0$ too high above the BB, the whole ray had to
be moved down. In the first step, the $r$-coordinate of the reverse ray was
retained, but its $t$ coordinate was lowered by $\Delta t_1 \df \Delta T \times
(1 + z_{f4})$, where $1 + z_{f4}$ is given by (\ref{8.4}) and $\Delta T$ is the
difference between $t(0)$ on IR 4 and the desired $t(0)$ on Ray 4. The
discrepancy decreased, but was still too large. So the next values of the
initial $t$ at $r_{\rm now4}$ were tested by trial and error, by adding
numerical coefficients to $\Delta t_1$. After a few corrections, the coincidence
shown in Fig. \ref{drawzbokudol} was achieved with $\Delta t_2 \approx -6.47358
\times 10^{-6}$ NTU; the initial point of the fine-tuned reverse ray is point B
in Fig. \ref{drawzbokugor}. Then, a future-directed axial ray was sent from
point B, and it intersected the $t = 0$ surface at point C in Fig.
\ref{drawzbokugor}. Actually, the ray again overshot $t = 0$ slightly, and the
coordinates of its endpoint are
\begin{eqnarray}\label{c.1}
t &=& 1.9143125092526522 \times 10^{-11}\ {\rm NTU}, \nonumber \\
r &=& 0.95585224106471711.
\end{eqnarray}
This became the initial point of the past-directed Ray 6, given by (\ref{9.1}).

%Real correction to initial t =   -6.4735809879400987E-004, calculated by
%downfromnow2.f90
%Ray 6 in the paper was calculated by the program downfromnow3.f90

\section{Remarks on numerical precision}\label{numprec}

To calculate the geodesics with a high precision, the numerical step in the
affine parameter, $\Delta \lambda$, should be as small as possible. But when it
is small, a single run of a numerical program lasts prohibitively long. A
compromise had to be struck. Between the LSH and $r = 0$ on Rays 1, 3 and 4 the
step was $\Delta \lambda = 10^{-9}$, in the same segment on Rays 2 and 5 it was
$\Delta \lambda = 10^{-6}$. On the segments of rays between $r = 0$ and the
present time $t = 0$, $\Delta \lambda$ was $10^{-8}$ on Ray 1 and $10^{-7}$ on
Rays 2, 3 and 5.

%The future Ray 4 in the paper is Ray 6 from the programs

Ray 4 was designed to be the representative one, so its segment between $r = 0$
and the present time was calculated with a higher precision. On it, the initial
$\Delta \lambda$ at $r = 0$ was $10^{-17}$, then it was multiplied by 100 at
each of $r = 0.0004$, 0.0005, 0.002, 0.005 and 0.07. The reason of this changing
$\Delta \lambda$ is that $1 + z = k^t = \dril t {\lambda}$, so where $z$ is
large (resp. small), $t$ changes by large (resp. small) increments of $\Delta t
= (1 + z) \Delta \lambda$. On a future-directed geodesic, the initial $z = 0$
and decreases along the way, so after a while $\Delta t$ becomes very small and
the calculation proceeds exceedingly slowly, requiring a huge number of
numerical steps.

The reverse occurs on past-directed geodesics: $\Delta \lambda$ must be
decreased along the way, or else increasing $z$ damages the precision. On Ray 6,
the initial $\Delta \lambda$ at $(t, r)_{\rm now6}$ was $10^{-9}$, then it was
divided by 100 at each of $r =$ 0.07, 0.005, 0.002, 0.0005 and 0.0004.

For the nonaxial rays grazing the QSS region, considered in Sec. \ref{angrad}, a
different scheme of changes in $\Delta \lambda$ had to be applied because they
leave the Friedmann region for only a brief time and cover larger segments of
$r$, so too high a precision would result in prohibitively long integration
times. On them, the initial $\Delta \lambda$ was $10^{-9}$, and it was divided
by 100 at each of $x =$ 0.17 and 0.002.

{\bf Acknowledgement} For some calculations, the computer algebra system
Ortocartan \cite{Kras2001,KrPe2000} was used.

\end{document}